\documentclass[10pt, preprint]{aastex}

\usepackage{graphics}
\usepackage[hyperfootnotes=false]{hyperref}
\usepackage{color} 
\usepackage{amsmath}

\def\beqa{\begin{eqnarray}}
\def\eeqa{\end{eqnarray}}
\def\beq{\begin{equation}}
\def\eeq{\end{equation}}

\newcommand\g{\gamma}

\def\g{\gamma}

\def\ln{{\rm ln}}

\usepackage{amssymb}
\usepackage{amsthm}
\usepackage{indentfirst}

\usepackage{epstopdf}
\usepackage{epsfig}
\DeclareMathOperator{\erf}{Erf}

\begin{document}


\title{Acceleration of Thermal Protons By  Generic Phenomenological
Mechanisms}


\author{Vah\'e Petrosian\altaffilmark{1,2} and Byungwoo Kang\altaffilmark{3}}
\affil{Department of Physics,
Stanford University, Stanford, CA 94305}
\altaffiltext{1}{Department of Physics,
Stanford University, Stanford, CA 94305. email: vahep@stanford.edu}
\altaffiltext{2}{Kavli Institute of Particle Astrophysics and Cosmology,
Stanford University, Stanford, CA, 94305.}
\altaffiltext{3}{Department of Physics,
Stanford University, Stanford, CA 94305. email: bkang@stanford.edu}

\shorttitle{Acceleration of Thermal Protons}
\shortauthors{Petrosian \&  Kang}



%\documentclass{article}
%\usepackage{epsfig}
%\setlength{\topmargin}{0.7in}
%\setlength{\headheight}{0pt}
%\setlength{\headsep}{0in}
%\setlength{\footskip}{.2in}
%%%\setlength{\footheight}{12pt}
%\setlength{\oddsidemargin}{-0.1in}
%\setlength{\evensidemargin}{-0.1in}
%\setlength{\textwidth}{6.5in}
%\setlength{\textheight}{8.8in}
%\parindent=0.15in
%\parskip=0.05in
%\pagenumbering{arabic}

%\def{\myemail}{vahep@stanford.edu}

\begin{abstract}

We investigate heating and acceleration of protons from a thermal gas with
a generic diffusion and acceleration model, and subject to Coulomb scattering
and energy loss, as was carried out in  Petrosian \& East
(2008) for electrons. As protons gain energy their loss to electrons becomes
important. Thus, we need to solve the coupled proton-electron kinetic
equation. We numerically  solve the coupled Fokker-Plank equations and
computes  the time evolution of the spectra of  both particles. We show
that this can lead to a quasi-thermal component plus a  high energy nonthermal
tail. We determine the evolution of  nonthermal tail and  the
quasi-thermal component. 
The results may be used to explore 
the possibility of inverse
bremsstrahlung radiation as a source of hard X-ray emissions from hot 
sources such as solar flares, accretion disk coronas and  the intracluster
medium of galaxy clusters.   We find that emergence of nonthermal protons
is accompanied by excessive heating of the entire plasma, unless the
turbulence needed for scattering and acceleration is steeper than
Kolmogorov and the
acceleration parameters,
the duration of the acceleration, and/or the initial distributions are
significantly fine-tuned. These results severely constraint the feasibility of
nonthermal inverse bremsstrahlung process producing hard X-ray emissions.
However the nonthermal tail may be the seed particles for further
re-acceleration
to relativistic energies, say by a shock.  In the Appendix we present some
tests of the integrity of the algorithm used and   present a new
formula for the energy loss rate due to inelastic proton-proton interactions.

\end{abstract}


\section{Introduction}
\label{sec:intro}

Most particle acceleration mechanisms invoked for astrophysical sources must
start with acceleration of low energy particles of the background
magnetized plasma with a thermal or Maxwellian distribution confined in a finite
volume with density $n$, temperature $T$ and magnetic field $B$. This initial
phase of acceleration can produce a distribution consisting of a
quasi-thermal component plus a nonthermal tail. Such distributions can be
approximated by the kappa
distribution (see e.g. Pierrad \& Lazar 2010) and can be the seeds in the
so-called
thermal leakage injection model in diffusive shock acceleration (see e.g.
Haysung et al. 2014).  The main goal of this
and an earlier paper [Petrosian \& East
2008 (PE08)] is to explore the possibility of producing a prominent nonthermal
tail by some generic phenomenological acceleration process. 
PE08 treated the generation of nonthermal tail in
the distribution of  electrons 
neutralized by a cold noninteracting proton population. Here we evaluate the
conditions required for the
generation of nonthermal proton spectra subject to similar energizing mechanism
using the coupled Fokker-Planck (FP) kinetic equations.

The nonthermal tails in kappa-like  electron distributions
could also be responsible for nonthermal
emission in the hard X-ray regime for hot plasmas ($T\sim
10^6 - 10^8$) such as in solar flares, as demonstrated in Hamilton \& Petrosian
(1992), and possibly in the hot
intracluster
medium (ICM) of some clusters of galaxies.
Many clusters show a significant signatures of nonthermal activity first
observed as
radio radiation  that is due to synchrotron emission by a population of
relativistic (Lorentz factor $\g\sim 10^4$) electrons in $B\sim \mu{\rm G}$
field. In addition, earlier observations by several hard X-ray
instruments indicated that, in addition to the well known thermal soft X-ray
radiation, some clusters show
excess radiation above 10's of keV that could not be fitted by a single
temperature thermal bremsstrahlung model. For a review of early observations see
Durret et al. (2008),
Rephaeli et al. (2008) and Ferrari et al. (2008).
Initially the excess radiation was assumed to  arise from
bremsstrahlung of a
nonthermal tail of electrons extending to $\sim 100$ keV. However, as stressed
by Petrosian (2001) and later rigorously proved by EP08, this scenario would
cause excessive heating of ICM. Subsequently it was suggested by Wolfe \&
Melia (2006) that a nonthermal tail in proton distributions extending to 100's
of MeV could also be a
source of  hard X-rays (10-100 keV) via inverse nonthermal bremsstrahlung
produced by their scattering of lower velocity thermal electrons.  This process
was also proposed as a possible source of hard X-rays in
solar flares (see review by Emslie \& Brown 1985) because of
the possible higher X-ray yield compared to electrons. Exploring
this possibility is subject of our paper. As we will show there are similar
difficulties with this scenario as well.

It should be noted, that in the ICM hard X-rays could also be produced
by the inverse Compton (IC) scattering of
CMB photons by the radio producing relativistic electrons. The difficulty with
this model is that it requires a lower magnetic field than indicated by
Faraday rotation observations (see e.g. Feretti et al. 1996; Clarke
et al. 2001) or that expected
from equipartition.
Reviews of these emission processes and possible acceleration mechanisms are
given in Petrosian et al. (2008) and Petrosian \& Bykov (2008).%
\footnote{All the review  articles cited above  can be found in 
the February 2008, Volume 134 issue of the Space Science Reviews, entitled 
{\it Clusters of Galaxies: Beyond the Thermal View}, by Kaastra et al. (2008).}
However, more recent observations have cast doubt on the reality of the claimed
hard X-ray excesses. In particular,  observation of Coma
by
{\it Suzaku} (Wik et al. 2009) and {\it NuStar} (Wik et al. 2014), and the
Bullet cluster (Astaldello et al. 2015) give only upper limits
consistent with the claimed relatively high magnetic fields from Faraday
rotation measurements. The
non-detection of many clusters by {\it Fermi}-LAT (Ackermann et  al. 2011 and
2014; Huber et al.  2011) supports this view and puts stringent constraints on
population of high energy cosmic rays that may be produced in the shocks
arising from  mergers during  large scale structure formations (see recent
review by Brunetti \& Jones 2014).

The crucial feature  in production of nonthermal electron or
proton tails is the
interplay between the particle-particle (and particle-external fields)
interactions that
causes energy loss and momentum diffusion,  and the acceleration or energizing
mechanism due to interaction of
particles with plasma  turbulence and/or converging flows (e.g. shocks). 
For nonrelativistic background plasma $(kT<10^9)$ K the energy loss is dominated
by Coulomb
collisions, which tend to
equilibrate electrons and protons distribution to a single temperature
Maxwellian. Acceleration rates lower than the Coulomb  rates (i.e.
with longer timescales) tend to only heat the plasma and higher acceleration
rates  lead to a runaway distributions.
Thus, for production of significant but not
excessive nonthermal population we need acceleration rates comparable to
the Coulomb rates, so that in EP08, dealing with acceleration of electrons
only, we used acceleration timescales  comparable to electron-electron (e-e)
Coulomb times. Since the latter are much shorter than that of electron-proton
(e-p) (and proton-proton (p-p)) collision times at all energies protons remained
decoupled and kept their initial distribution. As shown in PE08, production of
nonthermal electron tails required special conditions, at least for ICM
conditions where one needs to maintain a relatively constant temperatures of
$kT\sim 1-10$ keV over a Hubble time. This is not an issue for short lived solar
flares (Hamilton \& Petrosian 1992). 

The situation is more complicated for protons
because  the longer p-p collisional time
requires a longer acceleration time. Such a rate for electrons would lead to
their heating.  Therefore, we assume no, or much lower
rate of acceleration of electrons. This can come about for pure Alfv\'enic
turbulence which do not interact with low energy electrons. (Low energy
electrons interact with the whistler waves of the fast mode branch.) However,
this does not mean that electrons can be de-coupled from protons. This
is because proton-electron (p-e) and p-p collision times can be
comparable, so that some of the energy gained by protons can be transfered 
to electrons. Therefore, we
must treat the {\it coupled electron and proton kinetic equations}
simultaneously. 

In addition, the
energy dependence of the acceleration rate also plays an important role with
acceleration rates increasing with energy being more efficient in producing
nonthermal tails.

In \S 2, we derive and present 
coefficients for Coulomb interactions and stochastic acceleration to be used
in a Fokker-Plank (FP) kinetic  equation, and explain
our algorithm for solving the coupled FP equations of protons and
electrons. In \S 3, we apply the algorithm to the stochastic acceleration of
thermal background particles, and examine the time evolution of the proton and
electron distributions for various acceleration model parameters. In \S 4, we
summarize our results and discuss their implications. Some details of the
acceleration and a new formula for proton inelastic energy loss rate
due to pion production are presented in the Appendixes.

\section{Fokker-Planck Equations and its Coefficients}
\label{sec:FP}

In order to compute the time evolution of the proton and electron distributions
when the two species of particles interact not only internally but also mutually
via Coulomb collisions, we need to solve their coupled kinetic
equations. We use the FP kinetic equation using the following simplifying
assumptions. We assume a magnetized background plasma where particles are tied
to magnetic field lines and undergo momentum and pitch angle diffusion. We also
assume that the mean free path for scattering is much smaller than the size of
the
region which means that particle momentum distribution is isotropic. We further
assume  that the system is closed; that is, there is no escape or injection of
particles. In this case the pitch angle-averaged, spatially-integrated momentum
distribution $f(t,p)$ satisfies the simple equation 
$\partial f/\partial t=(1/p^2)\partial [p^2D_{pp}(\partial f/ \partial
p)]/\partial p$. For convenience we use the energy distribution defined as
$N(t,E)dE=4\pi p^2f(t,p)dp$ to obtain the following commonly used two forms of
the kinetic
equation:

\begin{equation}\label{FP}
\frac{\partial N}{\partial t}
 =  \frac{\partial}{\partial E}\left(D_{\rm EE}\frac{\partial N}{\partial
E}\right)
 - \frac{\partial [(A(E)N]}{\partial E}=\frac{\partial^2(D_{\rm
EE}N)}{\partial E^2}
-\frac{\partial [{\tilde A}(E)N]}{\partial E},
\end{equation}
where the energy diffusion coefficient  and 
the direct energy gain (or loss) rates are given as

\beq\label{coeffs}
D_{\rm EE}=\beta^2D_{pp}\,\,\,\,{\rm
and} \,\,\,\, A(E)=\zeta D_{\rm EE}/E \,\,\,\,{\rm
with} \,\,\,\, \zeta= \frac{2\gamma^2-1}{\gamma^2+\gamma}.
\eeq
Here $\beta=v/c$ is the particle velocity $v$ in unit of the speed of light
$c$ and the Lorentz factor $\gamma=(1-\beta^2)^{-1/2}$. As shown in Petrosian
\& Chen (2015) (see also Tsytovich, 1977) 
${\tilde A}(E)\equiv A(E)+d D_{\rm EE}/dE$
provides a more accurate
representation  of energy gain (or Coulomb energy loss) rate
than $A(E)$. 

In what
follows we will present  results from numerical solutions of the first form of
the above FP equation
including the effects of stochastic acceleration (SA) by turbulence and Coulomb
collision. 
We will assume a background plasma of constant density $n$ and  $B$ field (but
not temperature $T$), and a constant energy level and spectrum of turbulence
so that the transport coefficients $D_{EE}$ and $A$  are constant in
time.
Following PE08 we use the simple form
\begin{equation}
A_{\rm SA}(E) = E/[\tau_0(1+E_c/E)^q],
\end{equation}
or the characteristic acceleration time scale
\begin{equation}
\tau_{ac} \equiv E/A(E) = \tau_0(1+E_c/E)^q,
\label{taccel}
\end{equation}
described by the three parameters $q$, $\tau_0$, and $E_c$.
The thin black  (solid and dotted) curves in Figure \ref{ac-losst} show some
examples of
acceleration times used in the next section.

As described above, $\tau_0$, which is a measure of the acceleration time,
should be comparable to the relevant collision times and $E_c$ allows to
introduce a break in the energy dependence of the acceleration rate (making it
steeper or flatter at higher energies depending on the value of the index $q$).
The break will be important only for low values of $E_c$ and will have little
effect for $E_c\gg kT$. For SA by turbulence the index $q$ is related to the
spectral index $q'$ of turbulence. For relativistic energies and Alfv\'enic
turbulence $q=q'-2$. In the inertial range one expects a Kolmogorov or
Kraichnin spectrum with $q'=5/3$ or 3/2 so that  $q=-1/3$ and $-1/2$,
respectively. But the turbulence spectrum is expected to
be steeper in the damping range at smaller scales (large wave vectors) where $q$
may become positive. At lower energies this relation becomes more
complicated as interactions with many other modes become important (see
e.g Pryadko \& Petrosian 1997 and Petrosian \& Liu 2004). In what follows we
will use $q=1, 0, -1$ to include all possible cases.% 
\footnote{Although we limit our calculations to the SA by turbulence it
should be noted that our results are more general and will be similar to that
expected from acceleration by a shock. For example, the energy dependence of
acceleration  rate by a shock, which
depends on the pitch angle scattering, rather than momentum diffusion
coefficient (Petrosian 2012), will have similar energy dependence (see 
Fig. 1 of
Petrosian \& Chen 2014).}

\subsection{Coulomb Collisions}

In the cold target approximation, in which a test
particle
of charge $Z_ie$, mass $m_i$, and velocity $v_i$ collides with stationary target
particles of
charge $Z_je$, mass $m_j$ and number density $n_j$, the Coulomb energy diffusion
rate
is
negligible but the Coulomb energy loss rate 
\begin{equation}
\dot{E}^{\rm cold}_{ij} \equiv -{\tilde A}_{\rm Coul}= 4\pi \ln\Lambda Z_i^2
Z_j^2e^4n_j/(m_jv_i),
\end{equation}
where $\ln\Lambda\sim 20-40$ is the Coulomb logarithm.
In what follows we will be interested in electron and proton test particles
with $Z_i=1$. In most  astrophysical situations we are dealing with fully
ionized ions so the primary target particles are electrons and protons
so $Z_j=1$ also. Contributions to Coulomb losses from other ions are
negligible except from alpha particles with a relative number density
$n_\alpha/n_p\simeq 0.08$ and $Z_j=2$, which may contribute up to 20 \%.  Note
that this also means that $n_p/n_e=0.86$ and $n_\alpha/n_e=0.07$. If we define
$\tau_{\mathrm{Coul}} =
(4\pi r_0^2cn_e\ln \Lambda)^{-1}$, where $r_0$ is the classical electron radius,
then
\begin{equation}
\dot{E}^{\rm cold}_{ij} =
\left(\frac{m_ec^2}{\tau_{\mathrm{Coul}}\beta}\right)\left(\frac{m_e}{
m_j}\right)\left(\frac{ Z_j^2n_j}{n_e}\right). 
\end{equation}
The cold target approximation can be used for test particle energies $E_i\gg
E_j$ of the background plasma. For interaction with a thermal background plasma
with temperature $T$, when the test particle energy $E_i$ approaches $E_j\sim
kT$, we must use the hot plasma
energy exchange rates which yields a
finite energy diffusion rate and  can be an energy gain when 
$E_i<kT$. In PE08 we
used hot plasma rate equations for nonrelativistic electrons given e.g. by
Miller et al. (1996) and
Nayakshin \& Melia (1998). Here we are interested on interactions of both
electrons and protons. This requires generalization of the two functions 
$G(E,E')$ and $H(E,E')$ of PE08 that describe the loss and diffusion rates,
respectively, as follows:

Generalization of Equation  (7) in EP08  gives 
\begin{equation}\label{Gij}
G_{ij}(E_i,E_j) = 
\left\{
\begin{array}{lll}
-{\beta_j}^{-1}(m_e/m_i) & \mbox{for } \beta_i < \beta_j, \\
0 & \mbox{for } \beta_i = \beta_j, \\
{\beta_i}^{-1}(m_e/m_j) & \mbox{for } \beta_i > \beta_j.
\end{array}
\right.
\end{equation}
The meaning of $G_{ij}$ is clear; if the test particle is faster than the target
particle, it will lose energy to the target particle according to cold Coulomb
loss rate. Conversely, if it is slower than the target particle, it will gain
energy from the target particle. 
\noindent
From this we can calculate the energy exchange rate (positive for loss,
negative for gain) $\dot{E}_{ij}$  by integrating $G_{ij}$ over the
distribution
$N_j(E_j)$ of the target particles:
\begin{eqnarray} 
\dot{E}_{ij}(E_i) & = &  \frac{m_ec^2}{\tau_{\mathrm{Coul}}}\int^{\infty}_0
G_{ij}(E_i,E_j)N_j(E_j)dE_j\\ 
& = & \frac{m_ec^2}{\tau_{\mathrm{Coul}}\beta_i}\left(\int^{\frac{m_j}{m_i}E}
_0 N_j(E_j)dE_j - \int^{\infty}_{\frac{m_j}{m_i}E}
\frac{\beta_i}{\beta_j}N_j(E_j)dE_j\right).
\label{Edotint} 
\end{eqnarray}

In a similar manner, we derive the corresponding Coulomb diffusion coefficient
$D_{ij}$. Generalizing $H(E,E')$ given by Equation (10) of PE08, we
define $H_{ij}(E_i,E_j)$ by
\begin{equation}\label{Hij}
H_{ij}(E_i,E_j) =
\left\{
\begin{array}{lll}
{\beta_i}^2/(3\beta_j) & \mbox{for }  \beta_i < \beta_j, \\
\beta_i/3 & \mbox{for } \beta_i = \beta_j, \\
{\beta_j}^2/(3\beta_i) & \mbox{for } \beta_i > \beta_j.
\end{array}
\right.
\end{equation}
$D_{ij}$ is given by integrating $H_{ij}$ over the target particle
distribution: 
\begin{eqnarray}
D_{ij}(E_i)  & = & \frac{(m_ec^2)^2}{\tau_{\mathrm{Coul}}}\int^{\infty}_0
H_{ij}(E_i,E_j)N_j(E_j)dE_j\\
 & = & \frac{(m_ec^2)^2}{\tau_{\mathrm{Coul}}\beta_i}
\left(\int^{
\frac{m_j}{m_i}E}_0
\frac{\beta_j^2}{3}N_j(E_j)dE_j+\int^{\infty}_{\frac{m_j}{m_i}E}
\frac{\beta_i^3}{3\beta_j}N_j(E_j)dE_j\right).
\label{Dint}
\end{eqnarray}

Note that, while $G_{ij}$ is inversely proportional to the mass of the slower
particle $H_{ij}$ has no explicit dependence on the masses of any particles and
in particular for identical particles $H_{ii}=\beta_i/3$ is independent of
target particle mas or velocity.
%The mass
%independence
%of $H_{ij}$ may be guessed from the fact that, for Coulomb collisions of
%identical particles in the cold target approximation, the diffusion coefficient
%turns out to be independent of the mass of the colliding particles. Of course,
%this does not justify the mass independence of $H_{ij}$ when the two species
%$i$
%and $j$ have different masses. We will justify the mass independence of
%$H_{ij}$
%after we evaluate $\dot{E}_{ij}$ and $D_{ij}$ for Maxwellian distributions of
%target particles below.

When the number of particles in the nonthermal component of $N_j(E_j)$ is much
less than that of the thermal component we can approximate the distribution of
the target particles by
the nonrelativistic  Maxwellian form
\begin{equation}\label{Maxwell}
N_T(E) = \frac{2}{\sqrt{\pi}}(kT)^{-3/2}\sqrt{E}e^{-E/kT},
\end{equation}
which then yields the hot
Coulomb  coefficients 
\beq
\dot{E}^{\mathrm{hot}}_{ij}(E) \equiv -{\tilde A}_{\rm Coul}^{\rm hot}
=\frac{m_ec^2}{\tau_{\mathrm{Coul}}\beta}\left(\frac{m_e}{m_j}
\right)\left(\frac{
Z_j^2n_j}{n_e}\right)\left[\erf(\sqrt{x})-2\left(1+\frac{m_j}
{m_i}\right)\sqrt{\frac{x}{\pi}}e^{-x}\right],
\label{Edothot}
\eeq
and 
\beq
D^{\mathrm{hot}}_{ij}(E)
\frac{m_ec^2}{\tau_{\mathrm{Coul}}\beta}\left(\frac{m_e}{m_j}
\right)\left(\frac{
Z_j^2n_j}{n_e}\right)kT\left[\erf(\sqrt{x})-2\sqrt{\frac{x}{\pi
}}e^{-x}\right],
\label{Dhot}
\eeq
where $x = (E_i/kT_j)(m_j/m_i)$ and Erf stands for the Error Function.

\begin{figure}[!ht]
\begin{center}
\includegraphics[width=0.8\textwidth]{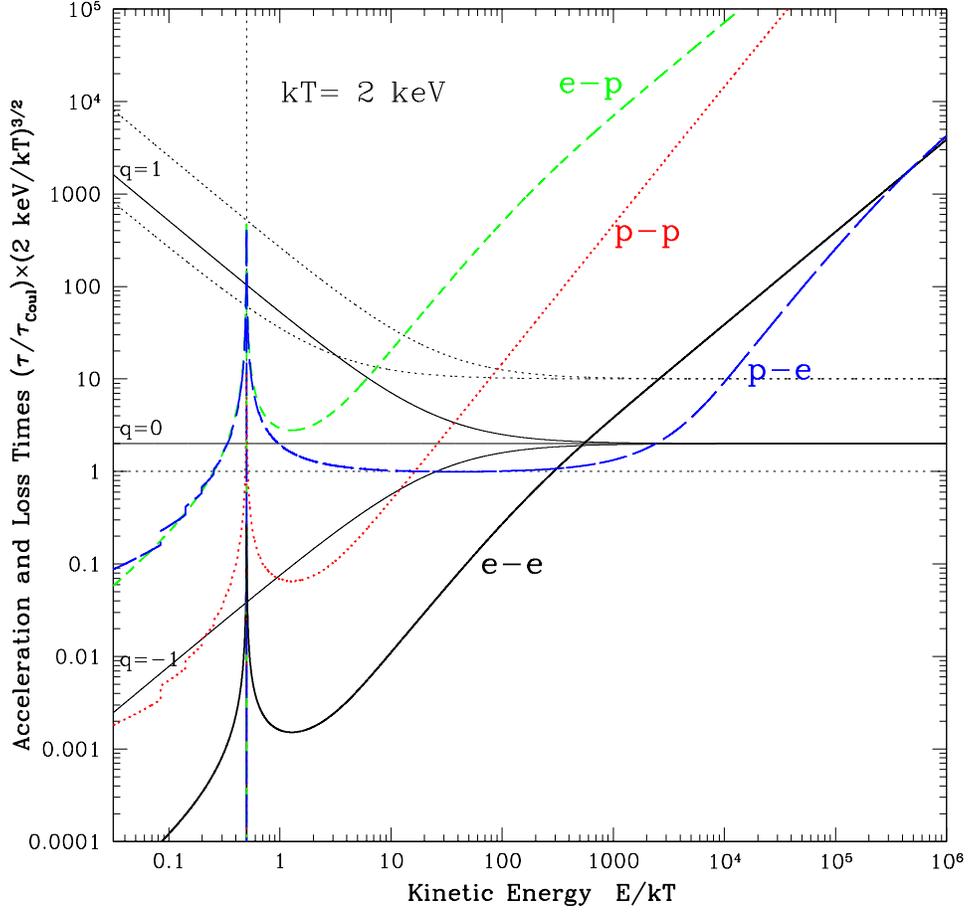}
\caption{Variation with test particle energy ($E$ in units of target particle
temperature $kT$) of the effective Coulomb loss timescales
$E_i/\dot{E}^{\mathrm{eff}}_{ij}$ (in
units of $\tau_{\rm Coul}$) 
for the four different pairs of the test and target particles; solid (black)
for e-e, dotted (red) for p-p, dashed (green) for e-p, and long-dash (blue)
for p-e. Effects of 8\% by number of He in a fully ionized plasma are included.
At $E/kT = 1/2$ the effective Coulomb loss
timescale
diverges (see Equation.($\ref{eq:Couleff}$)), and for  $E/kT<1/2$ the test
particle
actually gains energy (i.e. $\dot{E}_{\rm hot}^{\rm eff}<0$). The  lines
below this energy show the timescales using the absolute values of the loss
rate. Note that even
though we have used the non-relativistic approximation at low energies these
timescales give the correct results in the relativistic regime where the masses
of test  particles are irrelevant.
The thin solid (for $\tau_0=2\tau_{\rm Coul}$; $q=1, 0, -1$;  $E_c=25.5kT$) and
dotted (black) curves (for $\tau_0=10\tau_{\rm Coul}; q=1: E_c=2.55kT$ and
$E_c=25.5kT$) give several examples of the acceleration time according to
Equation (\ref{taccel}). Note that all temperatures refer to the
temperature of the target particles and assumed to be the same for electrons
and protons.}
\label{ac-losst}
\end{center}
\end{figure}

As mentioned above our numerical code solves the
first form of the FP equation,
given in Equation (\ref{FP}) which has the energy gain (loss) coefficient
$A={\tilde A}- d D_{\rm EE}/dE$. This means  that the Coulomb energy loss rate 
should also be transformed to $\dot{E}^{\mathrm{eff}}_{ij} = 
\dot{E}^{\mathrm{hot}}_{ij} +
\frac{dD^{\mathrm{hot}}_{ij}}{dE}$ or 
\begin{equation}
\dot{E}^{\mathrm{eff}}_{ij} =
\left(\frac{m_ec^2}{\tau_{\mathrm{Coul}}\beta}\right)\left(\frac{m_e}{m_j
}\right)\left(\frac{ Z_j^2n_j}{n_e}\right)\left(\erf(\sqrt{x})-2\sqrt
{\frac{x}{\pi}}e^{-x}\right)\left(1-\frac{kT_j}{2E}\right).
\label{eq:Couleff}
\end{equation} 
Note that at $E_i/kT_j = 1/2$ the effective Coulomb loss
timescale
diverges  and for $E_i/kT_j<1/2$ the rate $\dot{E}_{\rm hot}^{\rm eff}<0$ which
means that the test
particle
actually gains energy from the target particles. 
Figure \ref{ac-losst} shows the
effective Coulomb loss timescales, defined as $E_i/\dot{E}^{\mathrm{eff}}_{ij}$,
as a function of $E_i/kT_j$ for the four different pairs of the test and target
particles (e-e, p-p, e-p, p-e).%
\footnote{Inclusion of the effects of $\alpha$ particles with $n_\alpha=0.08n_p$
does not change $\tau_{ee}$ and $\tau_{pe}$ but increases   $\tau_{pp}$ and
$\tau_{pe}$  by 1.16. Adding losses to $\alpha$ particles reduces the two latter
times by 8\% (to 1.07) at relativistic energies and $\tau_{pp}$ by 10 to
30\% around $E\sim kT$. These effects are included in Figure  \ref{ac-losst}.}
There are several  features in these equation
that deserve some discussion.

\begin{enumerate}

\item The first feature is that, these rates  and timescales, although 
obtained using the
nonrelativistic Maxwellian distribution  approximation,  give the correct
(cold target) loss timescales for
 test particles with $E_i\gg kT_j$.

\item These forms of the Coulomb rates also satisfy the
time-independent FP equations  for both species of particles, when their
distributions are Maxwellian of the same temperature (see Appendix A)
and as shown in Appendix B they conserve energy.

\item As also can be deduced from the above equations the  shapes of the
timescales
shown in Figure \ref{ac-losst} are invariant 
and independent of $kT$ but they scale with the temperature of the target
particles as $(kT_j) ^{3/2}$ (see  Appendix C). Thus, although for the purpose
of comparison with the acceleration timescale we have used $kT=2$ keV, the loss
times as scaled in the vertical axis are valid for all temperatures.

\item Finally the relative values of the timescales play an important role. When
electrons are energized the acceleration timescale should be comparable to the
shortest e-e collision loss time. In this case the electrons gain energy but
share very little of it with protons since e-p timescale is $m_p/m_e=1836$
times longer. But as electrons are heated the e-e timescale increases as
$(kT_e)^{3/2}$ and when $kT_e\rightarrow 150\times kT_p$ some of the energy
goes into protons. This is why EP08, limiting their calculations to lower
temperatures, did not need to deal with a coupled kinetic equation.
On the other hand, when protons are energized, which is the case we are
considering here, the initial
phase is similar (p-p time is shorter than p-e) and only  protons
gain energy. However, as can be seen from Figure \ref{ac-losst}, once the
proton temperature increases by a factor less than ten the p-e interactions
become
important and some of the energy goes to electrons which are thermalized
quickly because of much shorter e-e timescale. For this reason, as stated
above, we need to solve the more complex coupled FP kinetic equations of
protons and electrons even though the acceleration mechanism is energizing
only the protons. The new algorithm used for this case is described next. 

\end{enumerate}

\subsection{Algorithm}

The algorithm we adapt iteratively solves the coupled FP equations of protons
and electrons
interacting via Coulomb collisions with only proton subject to SA. The
coupling arises from the fact that the 
energy lost to electrons by protons $\dot{Q}(E_e)\equiv \int_0^\infty
\dot{E}^{\rm hot}_{pe}(E_p, E_e)N(E_p)dE_p$ gives the  heating rate of 
electrons and vice versa.  At each time iteration step the algorithm uses the
Coulomb coefficients calculated based on particle distributions obtained in the
previous time step. PE08 fitted the distributions
in each step  to a Maxwellian with   ``effective temperatures'' and used 
the hot plasma Equations (\ref{Edothot}) and (\ref{Dhot}) to 
calculate the Coulomb collision
coefficients. This clearly is a reasonable approximation when the distributions
are nearly isothermal and saves considerable computing time. However, for
particle distributions
with strong nonthermal tails that are of  interest here this may not be
a good approximation. Moreover, when two species are involved 
it becomes more 
complicated to determine their respective effective temperatures.
{\it For these reasons, we use the more accurate but the  more time consuming 
algorithm whereby at each time step we calculate the  Coulomb
collision coefficients by integrating the more exact  Equations (\ref{Edotint})
and
(\ref{Dint})  over the particle distributions determined from
solution of the coupled FP equation at the previous time step.}
Several tests of the integrity of this algorithm are presented in Appendix C.

\section{Time Evolution of Proton and Electron Spectra}
\label{sec:Evol}

In this section we show the results from using the above relations and
algorithm to  investigate whether SA can generate a
significant  and discernible 
nonthermal proton tail without excessively heating the plasma. We assume that 
protons and electrons are both initially in equilibrium with  Maxwellian
distribution of $kT_0 = 2\;{\rm keV}$, and interact both internally and mutually
via Coulomb
collisions while protons  undergo SA by turbulence. As explained in
\S \ref{sec:FP}, the acceleration model is  specified by the choice
of parameters $q$, $E_c$, and
$\tau_0$. We give $E_c$  in units of $kT_0$, and $\tau_0$ is
expressed in
terms of $\tau_{\rm Coul}$. In addition $\tau_0$  the other
relevant times  $\tau_{pp}\propto kT_p^{1.5}$ and
$\mathrm{\tau_{pe}\propto kT_e^{1.5}}$ in
the plateau region; for the assumed initial temperature $kT_0=2$ keV 
the plateau is at $\sim \tau_{\rm Coul}$ (see Fig. \ref{ac-losst}). 

\begin{figure}[!ht]
\begin{center}
\includegraphics[width=0.48\textwidth, height=0.3\textheight]{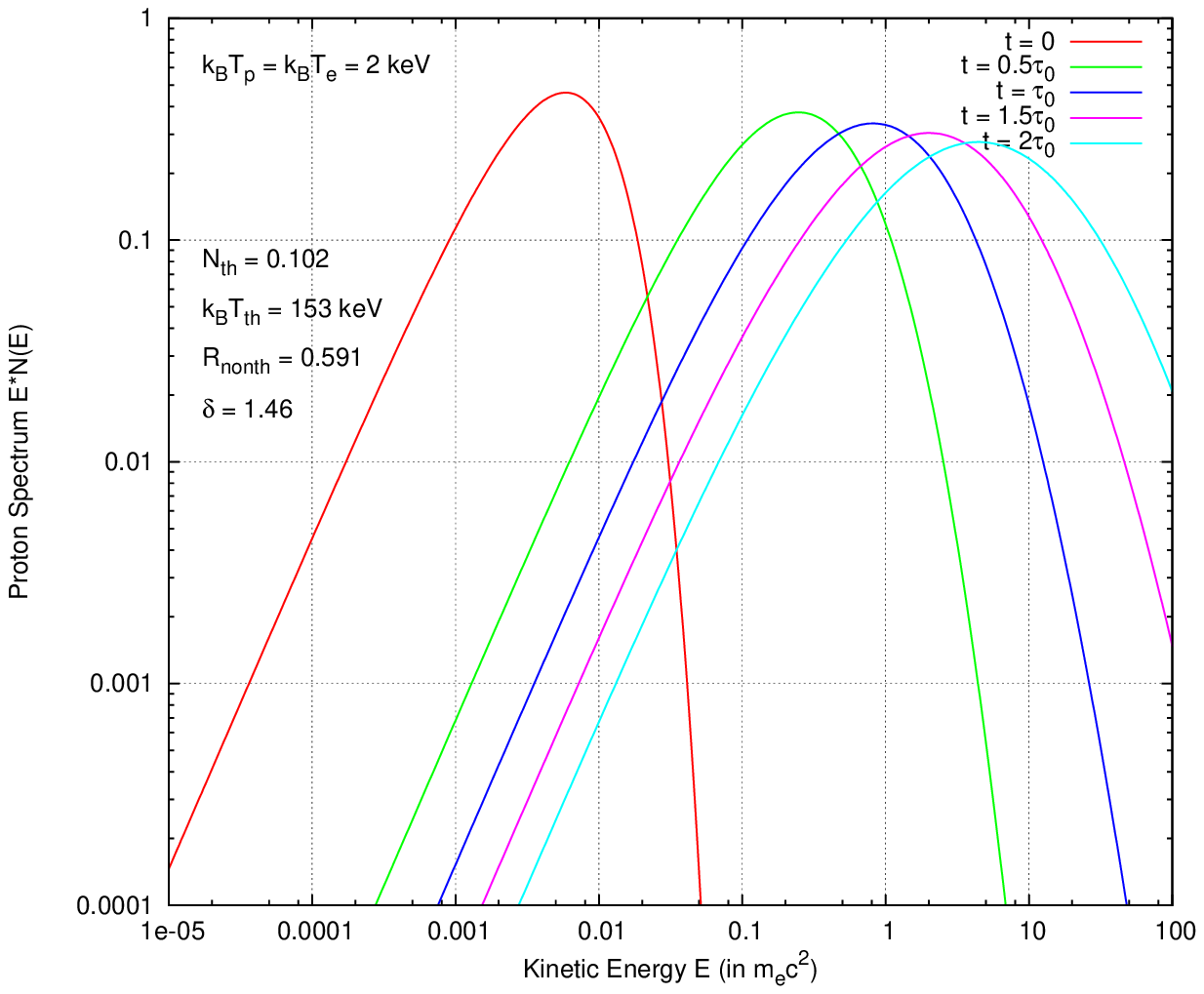}
\includegraphics[width=0.48\textwidth, height=0.3\textheight]{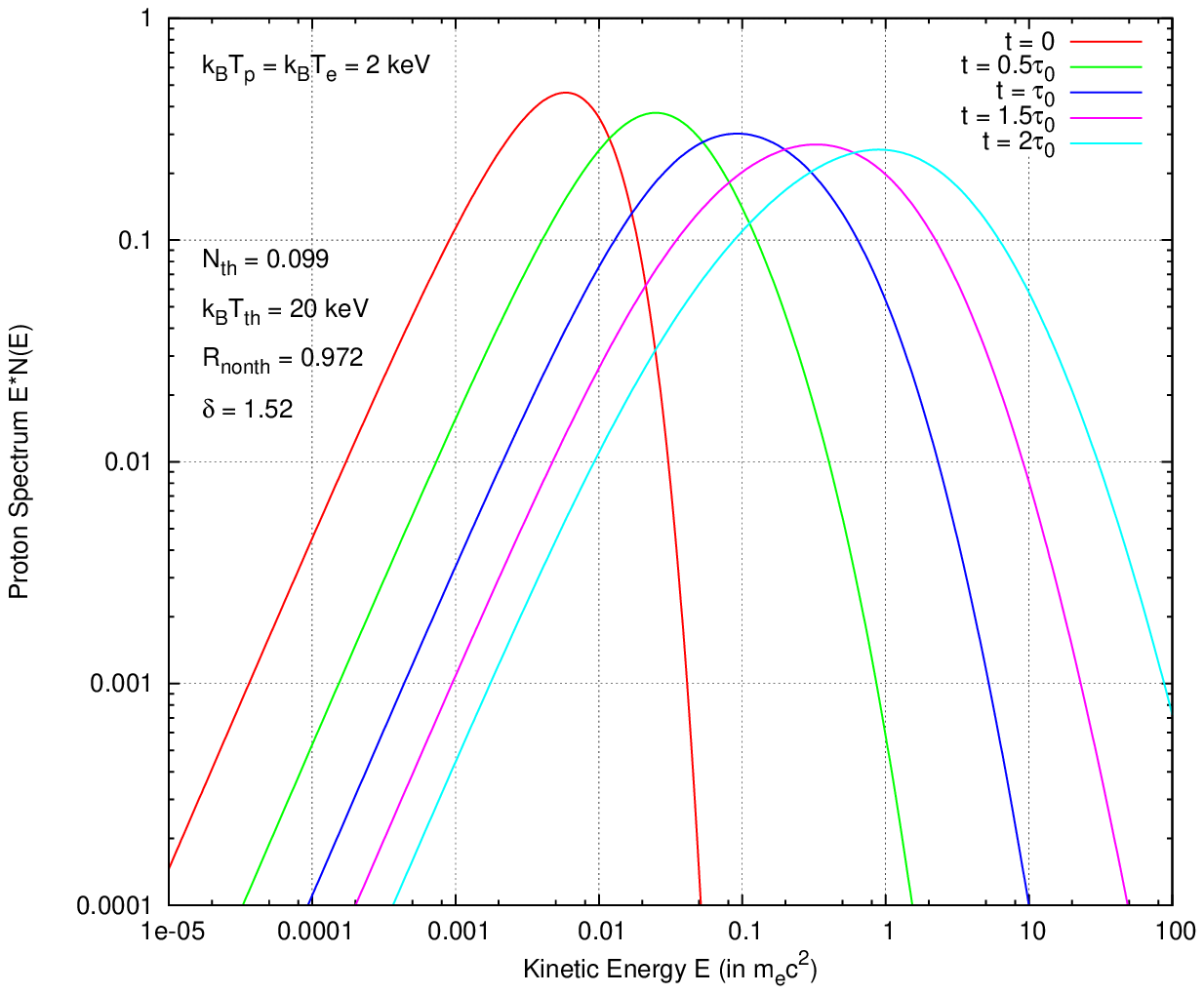}
\caption{Time evolution of the proton distributions in the presence of
turbulence that accelerates protons 
according to Equation  (\ref{taccel}), with $E_c =
25.5kT_0=0.1m_ec^2$,
$\tau_0=2.3\tau_{\rm Coul}$, 
and $q = -1$ (left) and $q=0$ (right).  Protons
and electrons are both assumed to be 
initially in the Maxwellian distribution with $kT_0 = 2\;{\rm keV}$ (left
curves), but shift to higher temperature monotonically in time as they
are accelerated at a constant rate and interact
both internally and with electrons 
via Coulomb collisions as described in section \ref{sec:FP}. The distributions
of protons begin to show more  deviation from pure Maxwellian  at later
times. Following EP08, in each figure, we
give the
fraction of number of protons in the thermal component 
$N_{\rm th}$, its temperature $kT_{\rm th}$, the ratio of the nonthermal energy
to total energy $R_{\rm nonth}$, (which should be 0.951 not 0.591 in the
left panel), 
and the approximate power-law index $\delta$ of the nonthermal component for the
spectrum at
the final time. Note that deviation from isothermal distributions  appear after
the proton temperature is increased by a factor of about 100.}
\label{fig:paccelq}
\end{center}
\end{figure}

\begin{figure}[!ht]
\begin{center}
\includegraphics[width=0.48\textwidth, height=0.3\textheight]{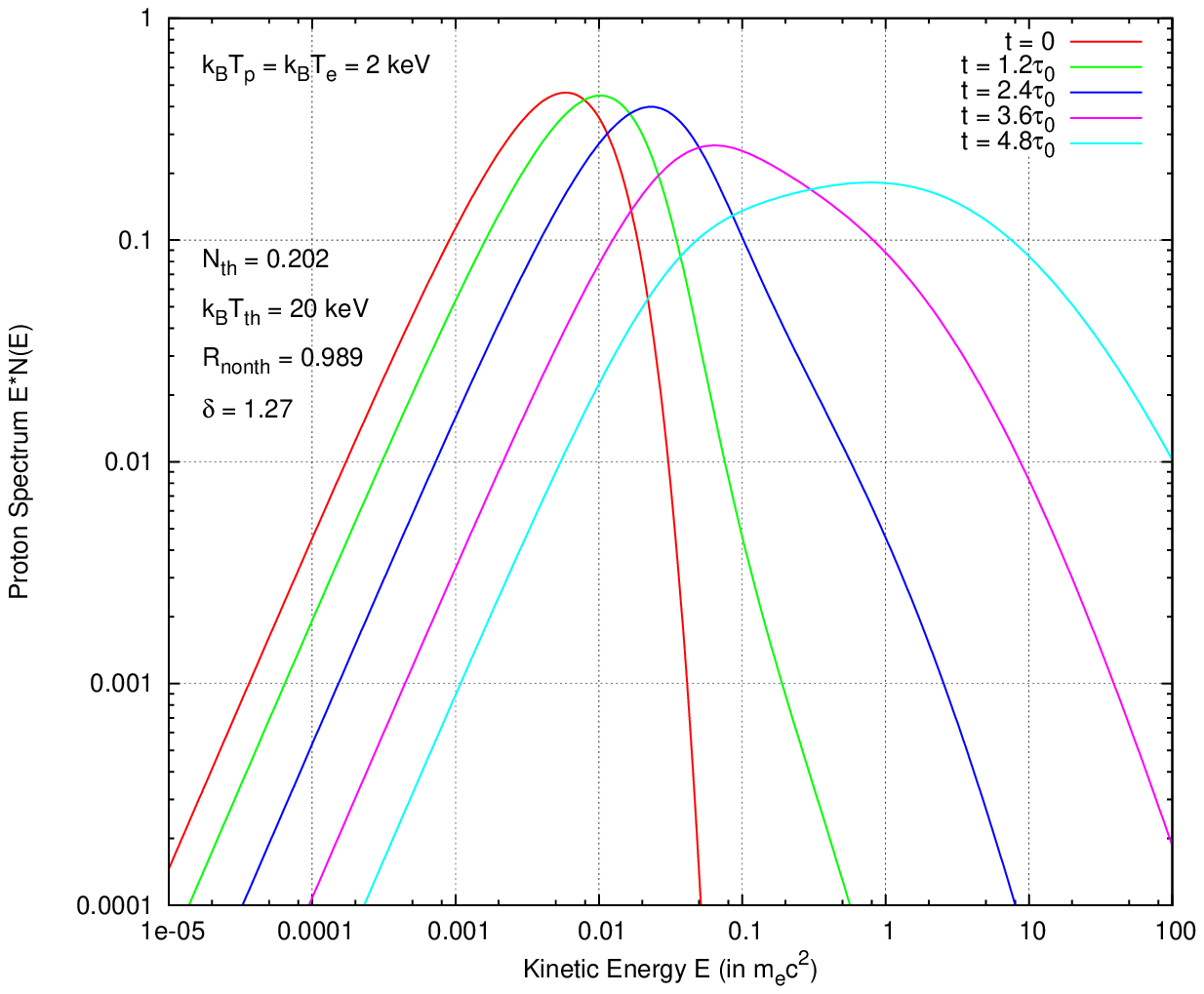}
\includegraphics[width=0.48\textwidth, height=0.3\textheight]{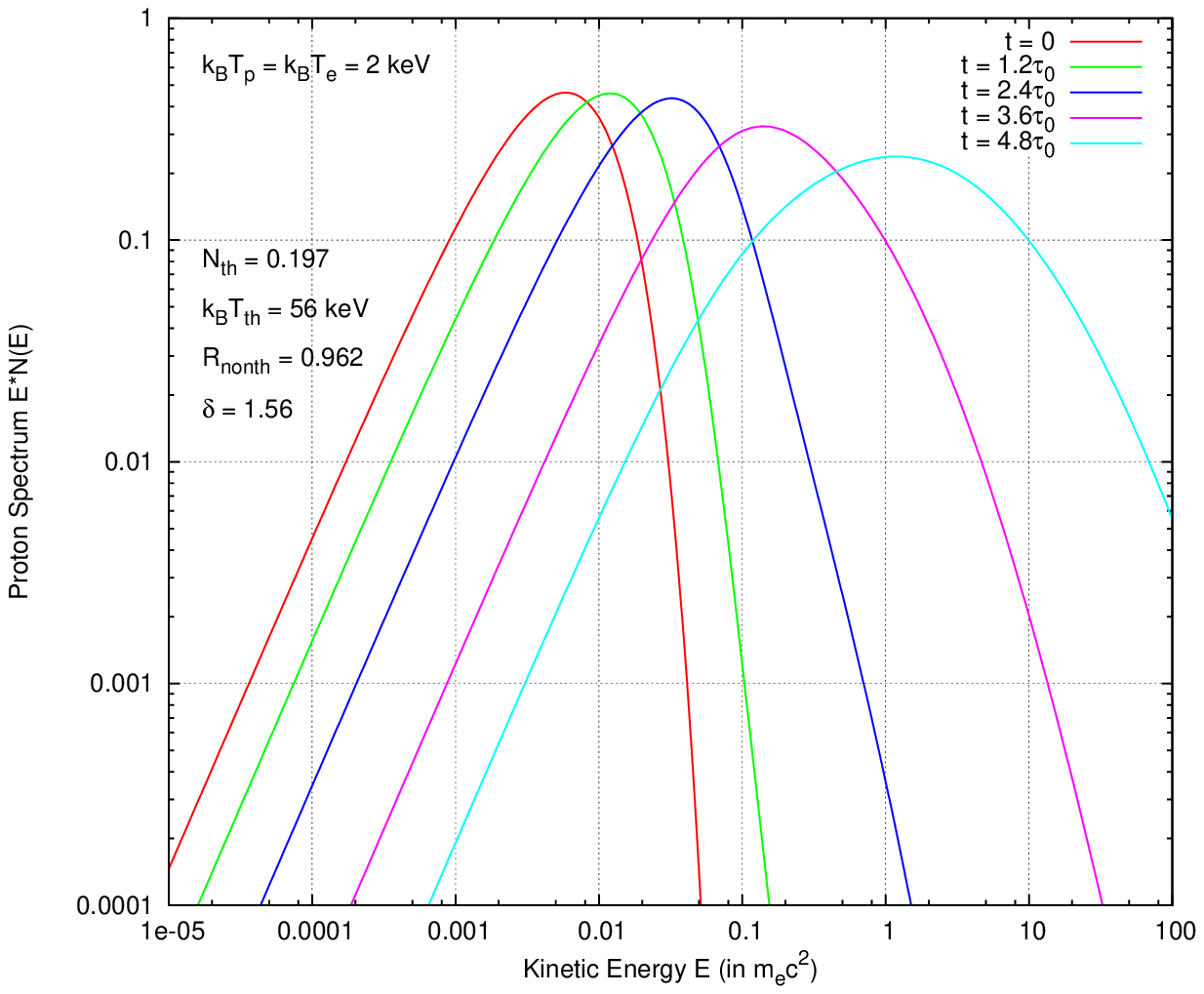}
\caption{Same as Figure \ref{fig:paccelq}, but with $q = 1$, $E_c = 25.5kT_0$,
and $\tau_0/\tau_{\rm Coul} = 2.35$ (left) and 13.5 (right).
Note that deviations from isothermal distribution start at lower temperatures
than for $q\leq 0$ models above and as expected lower $\tau_0$'s yield less
heating and stronger nonthermal component. }
\label{fig:pacceltau0}
\end{center}
\end{figure}

\begin{figure}[h!]
\begin{center}
\includegraphics[width=0.48\textwidth, height=0.3\textheight]{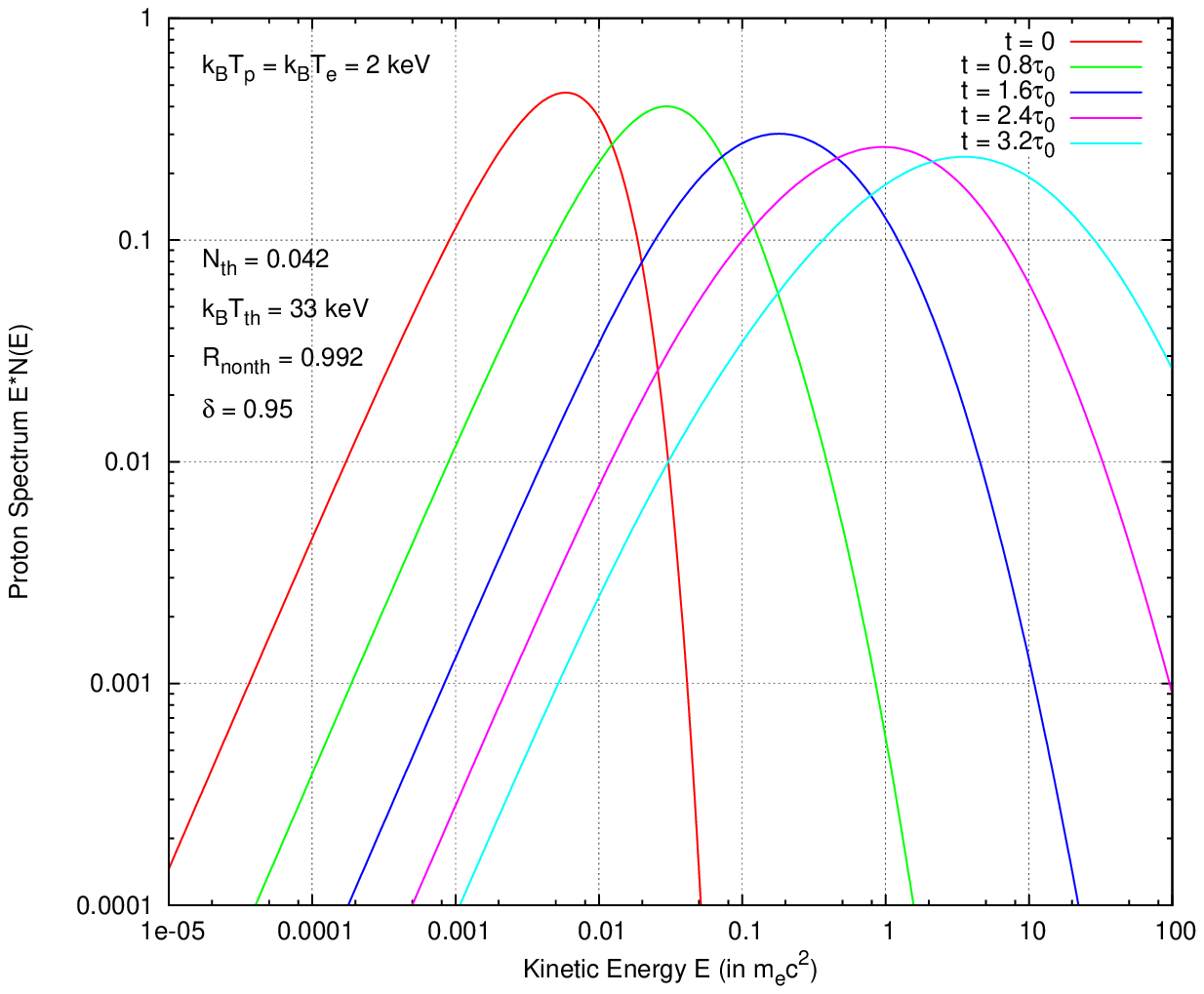}
\includegraphics[width=0.48\textwidth, height=0.3\textheight]{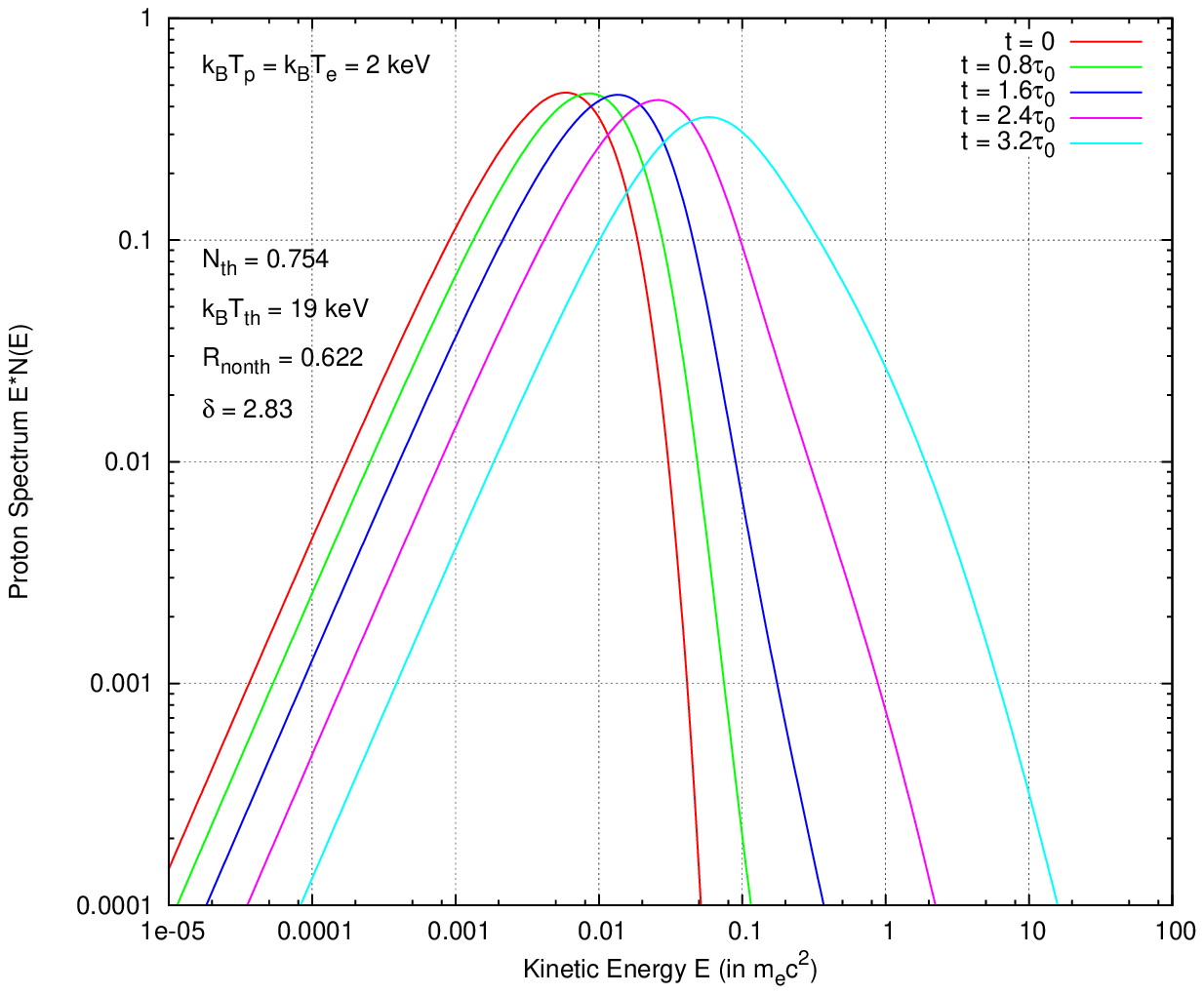}
\caption{Same as Figure \ref{fig:paccelq}, but with $q = 1$, $\tau_0
/\tau_{\rm Coul}=5.9$, and 
$E_c/kT_0=2.5$ (left) and 25.5 (right). As expected a higher value
of $E_c$ yields slower heating and weaker nonthermal component.}
\label{fig:paccelEc}
\end{center}
\end{figure}

In what follows we show time evolution of proton spectra, for several values of
the acceleration parameters $q, E_c/kT $ and $\tau_0$, at evenly
spaced time steps,
beginning with  initial Maxwellian distributions with $kT_0=2$ keV, which shifts
monotonically
to higher temperature and develops an increasing nonthermal component
resembling the so-called $\kappa$ distributions. Electron spectra, which tend
to be purely Maxwellian because of the short e-e collision times, are shown
only in two  interesting cases. These spectra are fit to a thermal
component of temperature $kT_{\rm th}$
plus an excess which we call the nonthermal component. We  give the the
temperature at the final step, and using the method described in the Appendix B
of PE08, we also give the fraction
of the number of protons in the thermal component $N_{\rm th}$,  the ratio
of the nonthermal  to total energy $R_{\rm nonth}$, and an approximate power-law
index $\delta$ of the nonthermal component (obtained by a fit to the excess
spectrum  at  energies one order of
magnitude above the energy at which it peaks). 

As in PE08, we also find that acceleration models with $q\leq 0$, 
which have stronger acceleration at lower energies where thermalization is fast,
lead primarily to heating and a small nonthermal component. Two examples
of these  are shown in
Figure \ref{fig:paccelq} for  models 
with $E_c = 25.5kT, \tau_0 = 2.3\tau_{\rm Coul}$ and $q = -1$ (left)
and $q=0$ (right). In both models (and others with similar values of $q$)
significant deviations from an isothermal distribution appears after the plasma
temperature is increased by several decades, which happens after 
$t\simeq 10\times \tau_{\rm Coul}\sim 10^ 8(10^{-3}{\rm cm}^{-3}/n)$ yr.

\begin{figure}[t!]
\begin{center}
\includegraphics[width=0.48\textwidth]{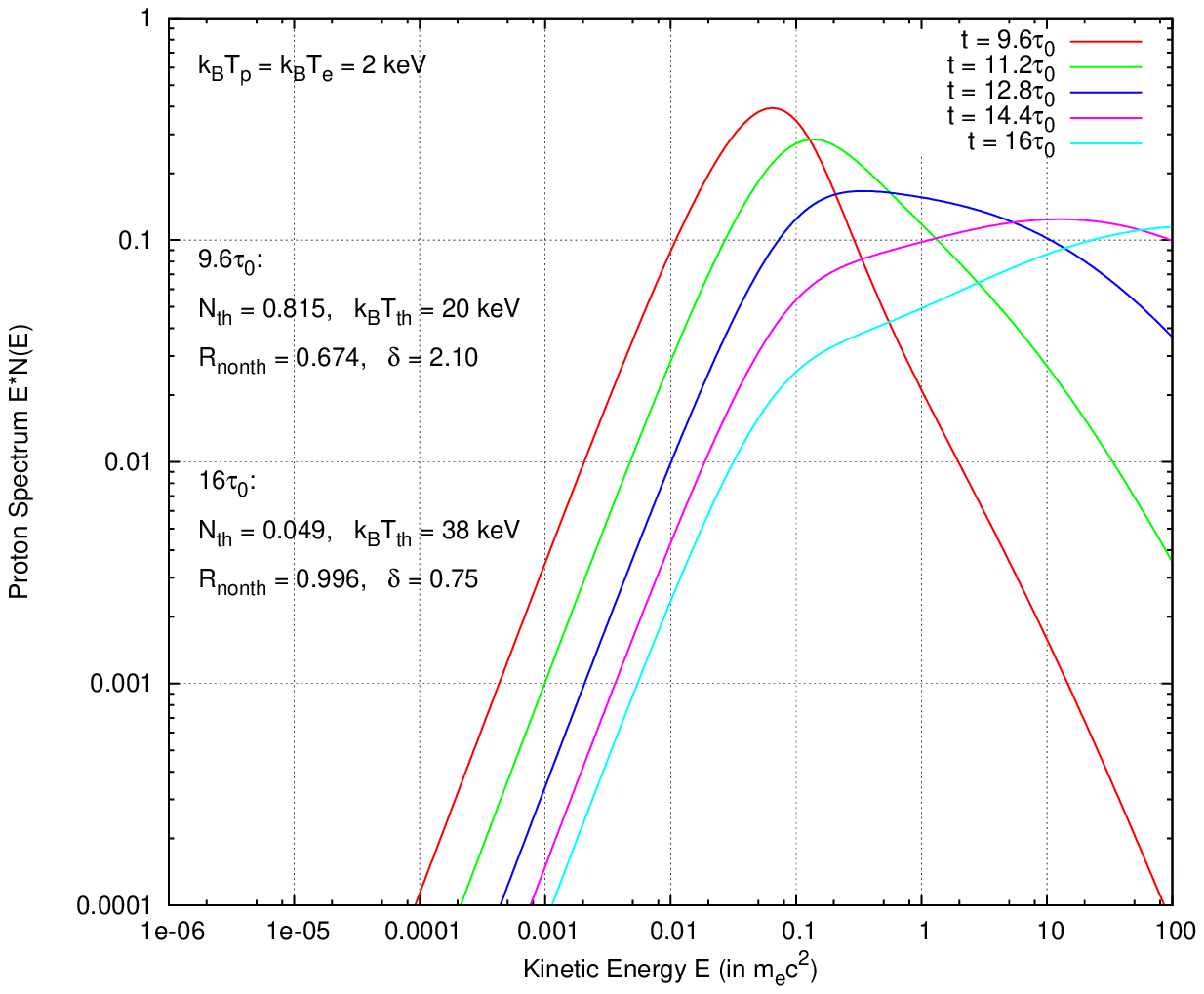}
\includegraphics[width=0.48\textwidth]{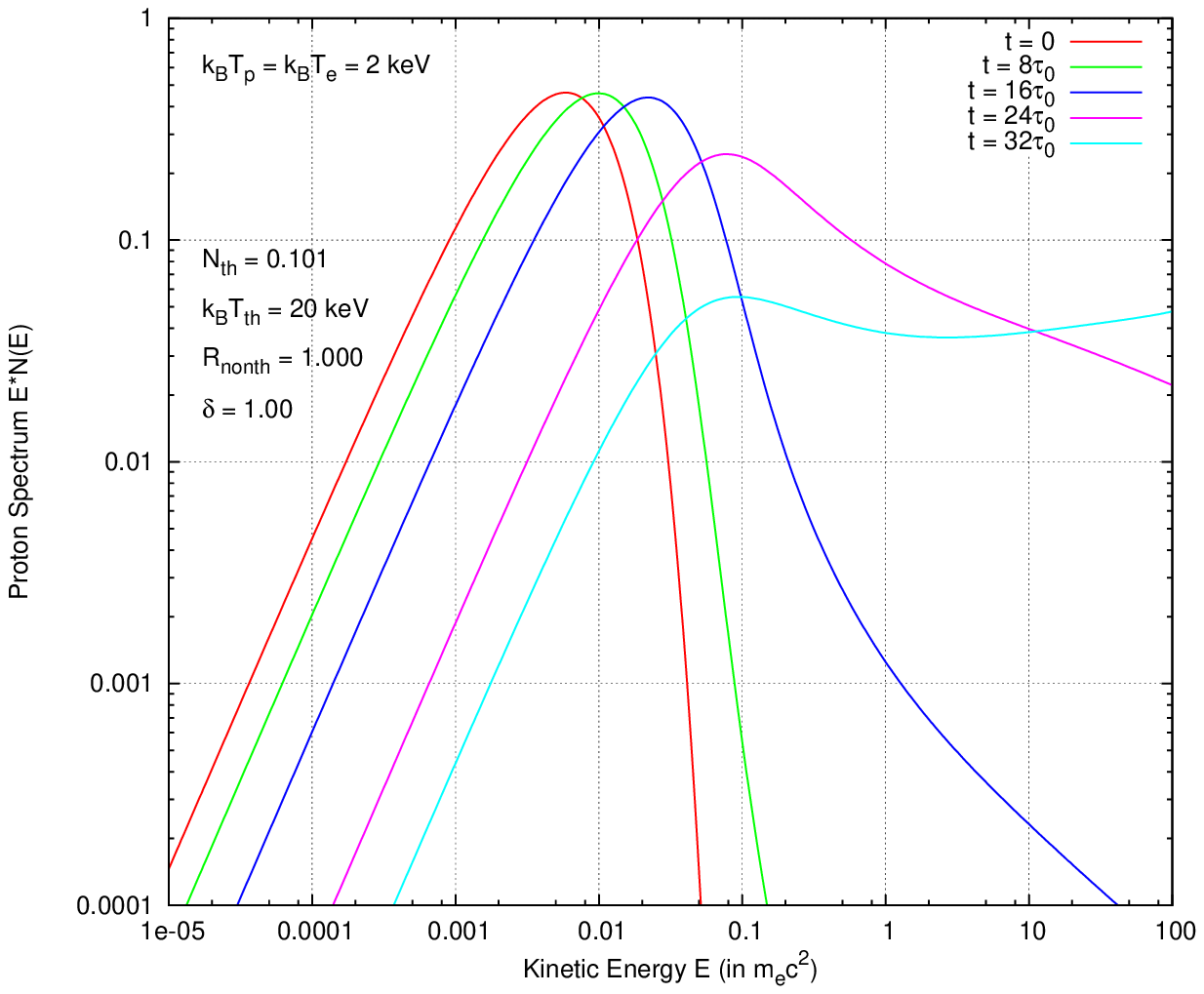}
\caption{Same as Figure \ref{fig:paccelq}, but with $q = 1$, $\tau_0 =
5.9\tau_{\rm Coul}$ and  $E_c=125kT_0$ (left)
and 
$\tau_0 = 1.2\tau_{\rm Coul}$, $E_c = 255kT_0$ (right).  For comparison
purpose,
for
the left plot  we
present $N_{\rm th}$, $kT_{\rm th}$, 
$R_{\rm nonth}$, and $\delta$ for both the earliest ($t = 9.6\tau_0$) and last
($t = 16\tau_0$).}
\label{fig:paccellate}
\end{center}
\end{figure}

\begin{figure}[!ht]
\begin{center}
\includegraphics[width=0.48\textwidth]{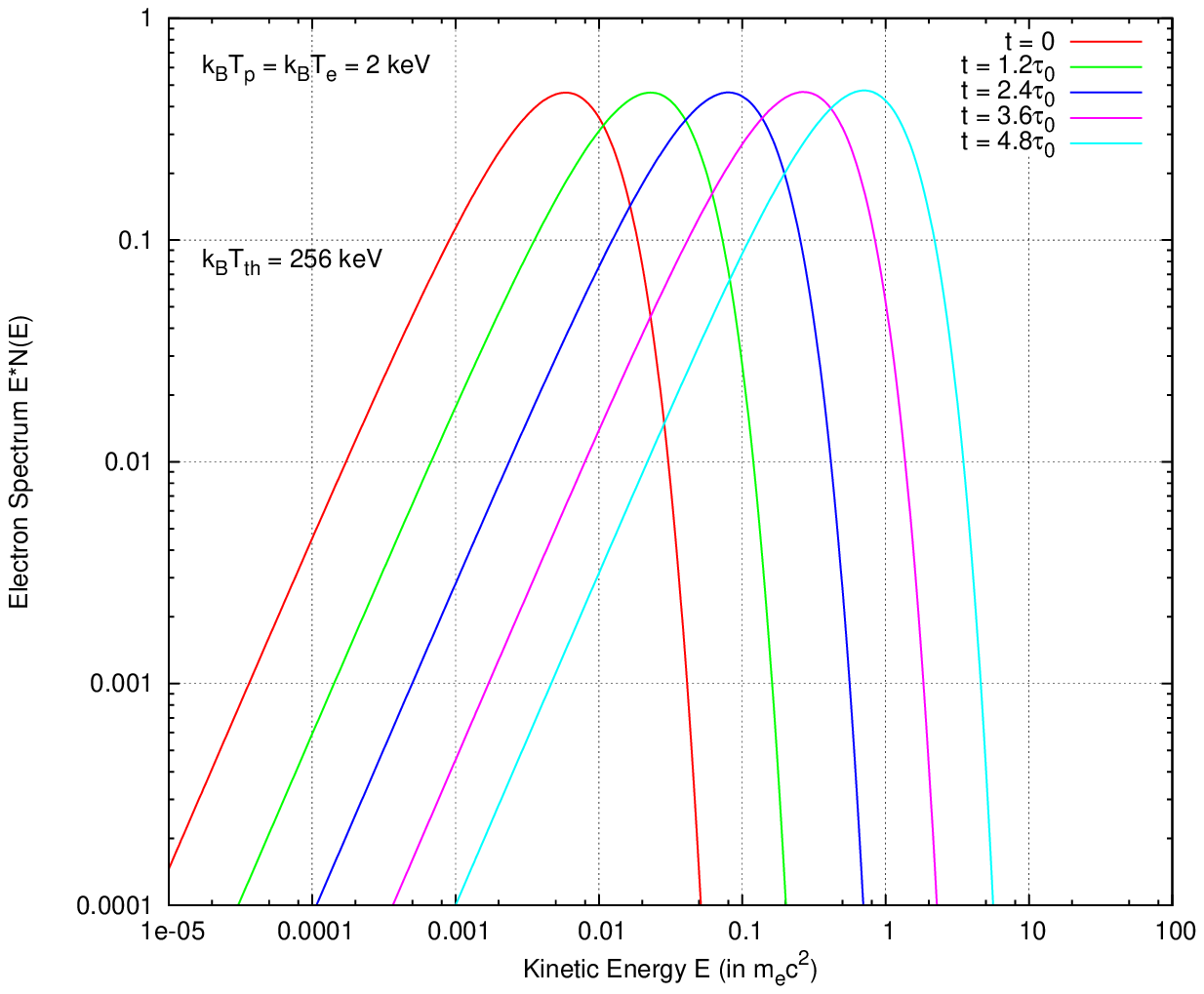}
\includegraphics[width=0.48\textwidth]{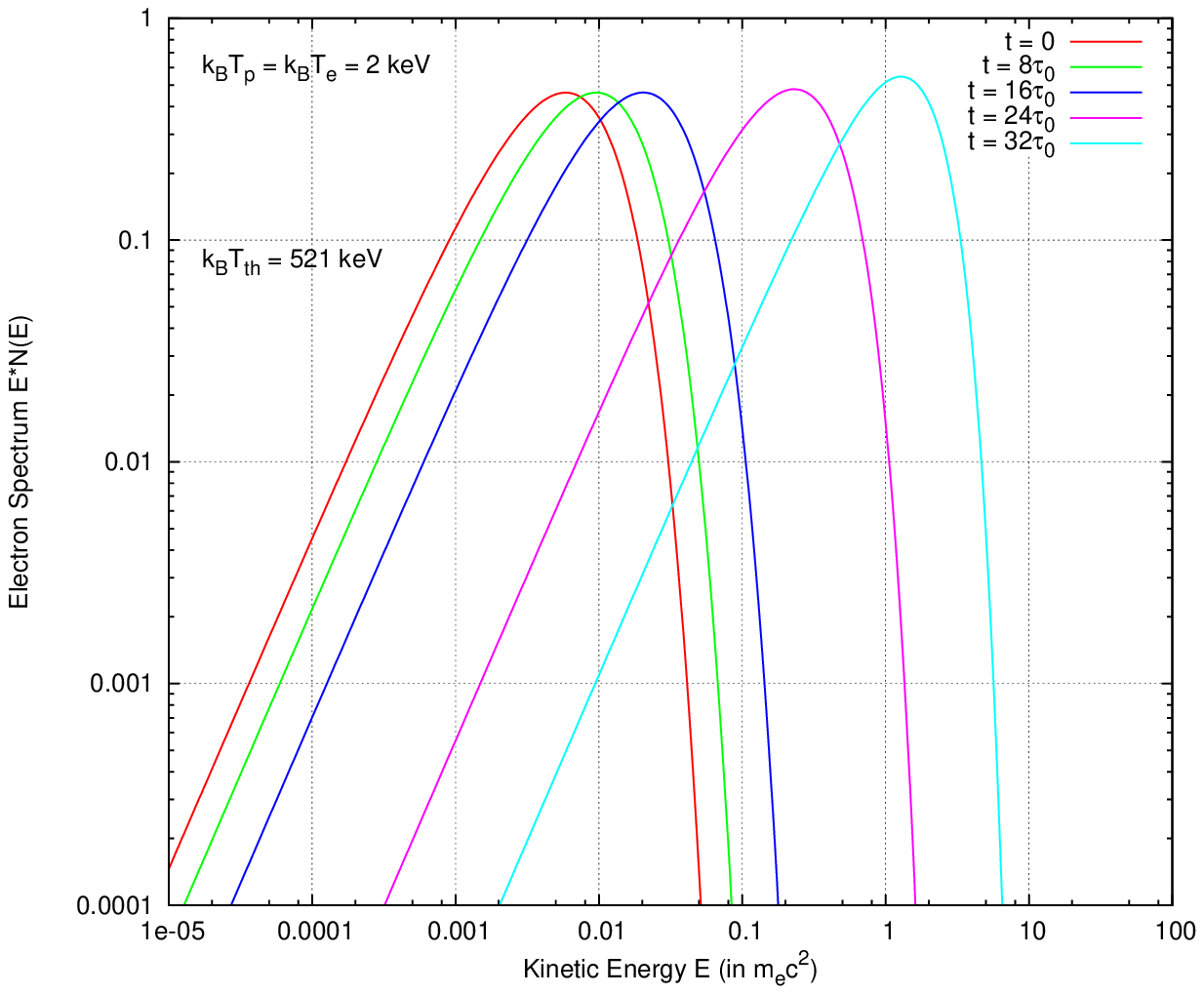}
\caption{Same as Figure \ref{fig:paccellate} but showing the time evolution of
the electron distributions. Note that because of much shorter
thermalization time of electrons the distributions remain thermal but with
increasing temperature.
The temperature of the final
electron distribution is indicated on each plot.}
\label{fig:paccelele}
\end{center}
\end{figure}

In what follows we  focus on $q = 1$ and  show how the
time evolution 
varies with $\tau_0$ in Figure \ref{fig:pacceltau0} and with $E_c$ in in Figure
\ref{fig:paccelEc}.  For the first set of  models with  $\tau_{ac}$ 
considerably longer
than thermalizing timescales of protons [see Equations (\ref{thermtaupp}) and
(\ref{thermtaupe})], most of the energy input from the turbulence goes into
heating rather than development of a distinct nonthermal tail, producing  broad
distributions similar to the $q = -1\;{\rm and}\;0$ models. {\it A
considerable nonthermal component emerges at late times 
after the temperature of the thermal component (or the average energy of the
distribution) becomes large enough so that the thermalization timescales become
considerably longer than $\tau_{ac}$}.%
\footnote{The fact that thermalization timescales are
increasing functions of energy also accounts for the late-time emergence of the
considerable nonthermal component in the $q = -1\;{\rm and}\;0$ models}.
The second set of models with a lower value of
$E_c$
exhibit faster increase of temperature  and a
more prominent  nonthermal component. In each
case,
the spectrum starts to show a significant nonthermal component only when the
proton temperature
approaches $E_c$. This behavior is due to the fact that
$\tau_{ac}\rightarrow \tau_0$ for $E \gg E_c$. On the other hand,  when
$\tau_0$ (or $\tau_{ac}$, to be more precise) is comparable
to or shorter than $\mathrm{\tau_{pe}}$, which is necessary for development of
a distinct
nonthermal tail, one can  avoid excessive heating.

Two examples that have large $E_c$ but $\tau_{0}$ comparable to or shorter than
$\mathrm{\tau_{pe}}$  are shown in Figure
\ref{fig:paccellate}. The  left panel shows the later  time evolution of
the proton spectrum
for the model $q = 1$, $\tau_0 = 5.8\mathrm{\tau_{\rm Coul}}$, and $E_c =
125kT_0$.
Although the spectrum at $t$ = $57\tau_{\rm Coul}$ exhibits a promising
nonthermal tail
and temperature that is only few times larger than $T_0$,  its shape
changes rapidly as a considerable fraction of the protons are accelerated beyond
$E_c$. By the time $t$ = $16\tau_0$, the spectrum is completely non-thermal
with a very small thermal component. On the right panel we show the  evolution
of
the proton spectrum for the model $q = 1$, $\tau_0 = 1.2\mathrm{\tau_{\rm
Coul}}$,
$E_c = 255kT_0$. In this case the spectrum remains nearly thermal until $t$ =
$8\tau_0$,
then begins to exhibit an appreciable nonthermal tail at $t$ =
$16\tau_0$, and  very quickly becomes strongly nonthermal afterwards. The
common feature of the time evolution for these two  models is that
the proton spectrum stays largely thermal before  $kT_p\sim E_c$ and then
gets
very rapidly accelerated once a considerable fraction of protons obtains kinetic
energy larger than $E_c$, leading to a runaway spectrum. 

These results
imply that, in order to get a nonthermal tail by an acceleration model
with $q = 1$, small $\tau_0$ and large $E_c$, a great deal of fine-tuning in the
duration of acceleration would be necessary. Also, we may
see from these results that $E_c$ cannot be too high compared to the desired
final temperature of the thermal component because the spectrum will start to
develop a considerable nonthermal tail only as it comes close to $E_c$. The
above conclusions will also hold for acceleration models with $q > 1$, whose
energy dependence at low-energy (i.e. $E < E_c$) is even steeper than $q = 1$
models.
              
As expected from the discussion in \S 2.2, the electron spectrum, 
in every case considered above, remains nearly thermal with
its temperature being usually smaller than
that of the proton thermal component during the early times. But once the
proton energy exceeds 20 keV p-e interactions become important and electrons
are quickly heated to proton temperatures. 
This is true regardless of how nonthermal the proton spectrum is. In
Figure \ref{fig:paccelele} we show the 
evolution of the electron spectra for for the same models shown in Figure
\ref{fig:paccellate}. As expected, in some cases and later times  (when
$\tau_{pe} < \tau_{pp}$) the
electron temperature could be much larger than that of the proton thermal
component especially when the proton spectrum has a prominent nonthermal
component. And in some cases the electrons are heated to relativistic
temperatures. In such a case
the electron bremsstrahlung will exceed the proton (inverse) bremsstrahlung.
This 
is another
difficulty with production of hard X-rays by a proton rather than an
electron nonthermal tail. 

{\it In summery, the above results strongly suggest that producing a significant
proton
nonthermal tail   requires a significant fine tuning of  the
acceleration
parameters, the duration of the acceleration, and/or the initial distributions.
As a result, for example, maintaining such a spectrum that could produce hard
X-rays in the ICM over a Hubble time would be difficult.}

\section{Summary and Discussion}

In this paper, we have explored the possibility of development of nonthermal
tails in the spectrum of protons starting from a background (nonrelativistic)
thermal plasma. We use  a generic phenomenological
energizing mechanism with a simple three parameter form for energy diffusion and
acceleration rates. As briefly described after Eq. (\ref{taccel}) in \S2,
these forms can simulate a broad range of  acceleration
processes. The energizing is opposed by Coulomb energy losses. Such a process
may be in operation in the initial phase of most acceleration mechanisms and may
be important in collisional plasmas like in solar and stellar flares and in the
ICM of galaxy clusters and could produce nonthermal bremsstrahlung radiation in
the hard X-ray regime in collisions with the background electrons. 

We derived the Coulomb energy-loss and diffusion
coefficients in a general setting of a ``hot plasma" where the distribution of
target particles and
the masses of the test and target particles are arbitrary. We present 
analytic formulas for  Maxwellian distribution that asymptotically approach
the cold target relations at high energies. We have generalized the
algorithm developed by PE08 to solve the coupled electron-proton kinetic
equations. We show that it satisfies the steady state equilibrium
distributions (Appendix A), and the energy conservation (Appendix B),
In Appendix C we test our algorithm by demonstrating that in absence of
acceleration the particle distributions relax to Maxwellian within the expected
timescales starting from several non-equilibrium initial conditions. In
Appendix D we discuss the inelastic energy loss rates and present a new
analytic formula for proton energy loss rate due to pion production.

Using the algorithm, we  obtain the time evolution of the proton and electron
distributions.  Because of shorter collision time of electrons the
production of nonthermal electron tails require a higher rate of
acceleration and shorter acceleration time than those used here for the
protons. {\it Thus, in our simulations we include only acceleration of protons
which may be the case in the presence of low frequency Alfv\'enic turbulence.
The
electrons gain energy from protons but thermalize quickly.} The resulting proton
spectra are decomposed into thermal
and nonthermal components according to the fitting methods of PE08. Our
results can be summarized as follows:

\begin{enumerate}

\item For $q = -1\;{\rm and }\;0$, even when acceleration time $\tau_0$ is
sufficiently small so
that the energizing rate is comparable to or faster than the thermalization
rate, an initially Maxwellian proton spectrum evolves into a hotter one and
broader than thermal distribution with little sign of a
distinct nonthermal tail. This is because for $q \leq 0$ the energizing rate is
smaller at higher energies and not efficient in acceleration into a
high-energy
tail.  This suggests that $q$ must be positive for production of a distinct
nonthermal tail and a mild temperature increase. This means we need
turbulence with spectrum considerably steeper than Kolmogorov.

\item We show that for $q = 1$ the temperature increases  more
slowly as
long as most of the particles have kinetic energy below $E_c$. In addition, a
significant nonthermal tail may appear if $\tau_0$ is comparable to or smaller
than the proton thermalization time.
For such small values of $\tau_0$, however, once a considerable fraction of the
protons have reached $E_c$, they rapidly evolves into a completely nonthermal,
runaway distribution. This will be more pronounced for even higher values of
$q$, where acceleration rate increases more rapidly with energy. 

\item The electron spectrum remains nearly thermal in every case considered in
this paper. For cases where the proton distribution is nearly Maxwellian the
electron temperature remains somewhat below that of the  proton. But when the
proton distribution contains a significant or a larger nonthermal component
(and $\tau_{pe}<\tau_{pp}$ for most protons) the
electron temperature can reach the average energy of the protons and 
be much higher than that of the proton thermal component.         

\end{enumerate}

{\it In summary, the above results shows that a great deal of fine-tuning and a steeper than Kolmogorov spectrum of turbulence is
necessary
for creating a significant nonthermal proton tail without excessive heating of
protons and/or electrons even when only protons are energized.} A corollary of
this is that, contrary to some claims (Wolfe \& Melia 2006: Boldt \&
Serlemitsos 1969), nonthermal inverse bremsstrahlung cannot be a significant
contributor to hard X-ray emissions (see  also Emslie \& Brown 1985).

\section*{Acknowledgements}
We thank Qingrong Chen for his extensive help in developing the numerical code.
This work was supported by Stanford University VPUE Major Grant No. 4842.

\appendix

\section{Proof of Steady State Maxwellian}

When two species of particles denoted by $i$ and $j$ are in thermal equilibrium,
the time-independent FP equation ( $\partial N/\partial t=0$) for each of the
two of the species, say $i$, is
written as  
\begin{equation}
0 = \frac{\partial^2}{\partial
E^2}[(D^{\mathrm{hot}}_{ii}(E)+D^{\mathrm{hot}}_{ij}(E))N_i(E)] +
\frac{\partial}{\partial
E}[(\dot{E}^{\mathrm{hot}}_{ii}(E)+\dot{E}^{\mathrm{hot}}_{ij}(E))N_i(E)],  
\end{equation}
where $N_i(E)$ is the (Maxwellian) distribution of the species $i$ and all the
Coulomb coefficients are evaluated at the same temperature as that of $N_i(E)$.
Since the $N_i(E)$ must be time-independent even in the absence of Coulomb
interactions with the other species of particles, i.e.,
\begin{equation}
0 = \frac{\partial^2}{\partial E^2}[D^{\mathrm{hot}}_{ii}(E)N_i(E)] +
\frac{\partial}{\partial E}[\dot{E}^{\mathrm{hot}}_{ii}(E)N_i(E)],  
\end{equation}
it follows that, in particular,
\begin{equation}
0 = \frac{\partial^2}{\partial E^2}[D^{\mathrm{hot}}_{ij}(E)N_i(E)] +
\frac{\partial}{\partial E}[\dot{E}^{\mathrm{hot}}_{ij}(E)N_i(E)].  
\end{equation}
Given $N_i(E)$ and $\dot{E}^{\mathrm{hot}}_{ij}$ we can integrate this equation
to uniquely determine
$D^{\mathrm{hot}}_{ij}(E)=\int_E^\infty\dot{E'}^{\mathrm{hot}}_{ij}
(E')N_i(E')dE'/N_i(E)$. The diffusion coefficient uniquely determined in this
way
agrees exactly with the ones presented in \S 2, which justifies the
assumptions used in their derivations. 

\section{Energy Conservation} 

Here, we show that, for arbitrary particle distributions and FP
coefficients are given by Equation  (\ref{Edotint}) and (\ref{Dint}),  our
algorithm
described in Section 2 satisfies the total energy conservation. Let $E_{tot}$ be
the total energy of the two species of particles.
Then, from their FP equations with no-flux boundary condition (see
Equation  (3) of Park \& Petrosian 1995), we find
\begin{equation}
\frac{\partial E_{tot}}{\partial t} = -(A + B + C + D),
\end{equation}
where
\begin{equation}
A =  \int^{\infty}_0 \int^{\infty}_0 N_i(E)G_{ii}(E,E')N_i(E')dEdE',
\end{equation}
\begin{equation}
B =  \int^{\infty}_0 \int^{\infty}_0 N_i(E)G_{ij}(E,E')N_j(E')dEdE',
\end{equation}
\begin{equation}
C =  \int^{\infty}_0 \int^{\infty}_0 N_j(E)G_{ji}(E,E')N_i(E')dEdE',
\end{equation}
\begin{equation}
D =  \int^{\infty}_0 \int^{\infty}_0 N_j(E)G_{jj}(E,E')N_j(E')dEdE'.
\end{equation}

A and B separately vanish because of  anti-symmetry of $G_{ii}$ and $G_{jj}$,
and because
$G_{ij}(E,E') = -G_{ji}(E',E)$, it follows that $B+C = 0$. This proves the
claim. 

\section{Thermalization Tests}

As a test of our algorithm, we consider the thermalization of the proton and
electron energy distributions that are initially not in thermal equilibrium.
The total particle numbers of the protons
and electrons are assumed to be the same and normalized to 1. The proton and
electron distributions are expected to converge to Maxwellian distributions of
the same temperature over certain thermalization timescales, conserving the
total energy at each time. Based on Equation  (\ref{Edothot}), we may define
four
thermalization timescales as
\begin{eqnarray}
\tau_{ee} (kT_e) & = &
\tau_{\mathrm{Coul}}\left(\frac{kT_e}{m_ec^2}\right)^{3/2} \sim
\left|\frac{kT_e}{\dot{E}^{\mathrm{hot}}_{ee}(kT_e)}\right|\label{thermtauee},\\
\tau_{pp} (kT_p) & = &
\tau_{\mathrm{Coul}}\sqrt{\frac{m_p}{m_e}}\left(\frac{kT_p}{m_ec^2}\right)^{3/2}
\sim
\left|\frac{kT_p}{\dot{E}^{\mathrm{hot}}_{pp}(kT_p)}\right|,\label{thermtaupp}\\
\tau_{ep} (kT_p,kT_e) & = &
\tau_{\mathrm{Coul}}\left(\frac{m_p}{m_e}\right)\left(\frac{kT_e}{m_ec^2}
\right)^{3/2} \sim
\left|\frac{kT_e}{\dot{E}^{\mathrm{hot}}_{ep}(kT_e)}\right|,\label{thermtauep}\\
\tau_{pe} (kT_p,kT_e) & = &
\tau_{\mathrm{Coul}}\left(\frac{m_p}{m_e}\right)\left(\frac{kT_p}{m_ec^2}
\right)^{3/2} \sim
\left|\frac{kT_p}{\dot{E}^{\mathrm{hot}}_{pe}(kT_p)}\right|,\label{thermtaupe}
\end{eqnarray}
where $kT_p$ and $kT_e$ are defined as $2/3$ of the total energy of the
initial distributions.%
\footnote{In the last expressions numerical factors of  order unity are omitted
and  for $\tau_{ep}$ and $\tau_{pe}$, appropriate asymptotic forms
of the function $\erf(\sqrt{x})-2\left(1+m_j/m_i\right)\sqrt{x/\pi}e^{-x}$  are
used.} Interestingly as can be seen from Figure 1, for $T_e\sim T_p$ the
interspecies interaction times 
$\tau_{ep}$ and $\tau_{pe}$ are almost equal and are  $\sqrt {m_p/m_e}$
and 
${m_p/m_e}$ times longer than $\tau_{pp}$ and $\tau_{ee}$, respectively.
Thus thermalization is governed by the former timescales.

\begin{figure}[!ht]
\begin{center}
\includegraphics[width=0.48\textwidth, height=0.3\textheight]{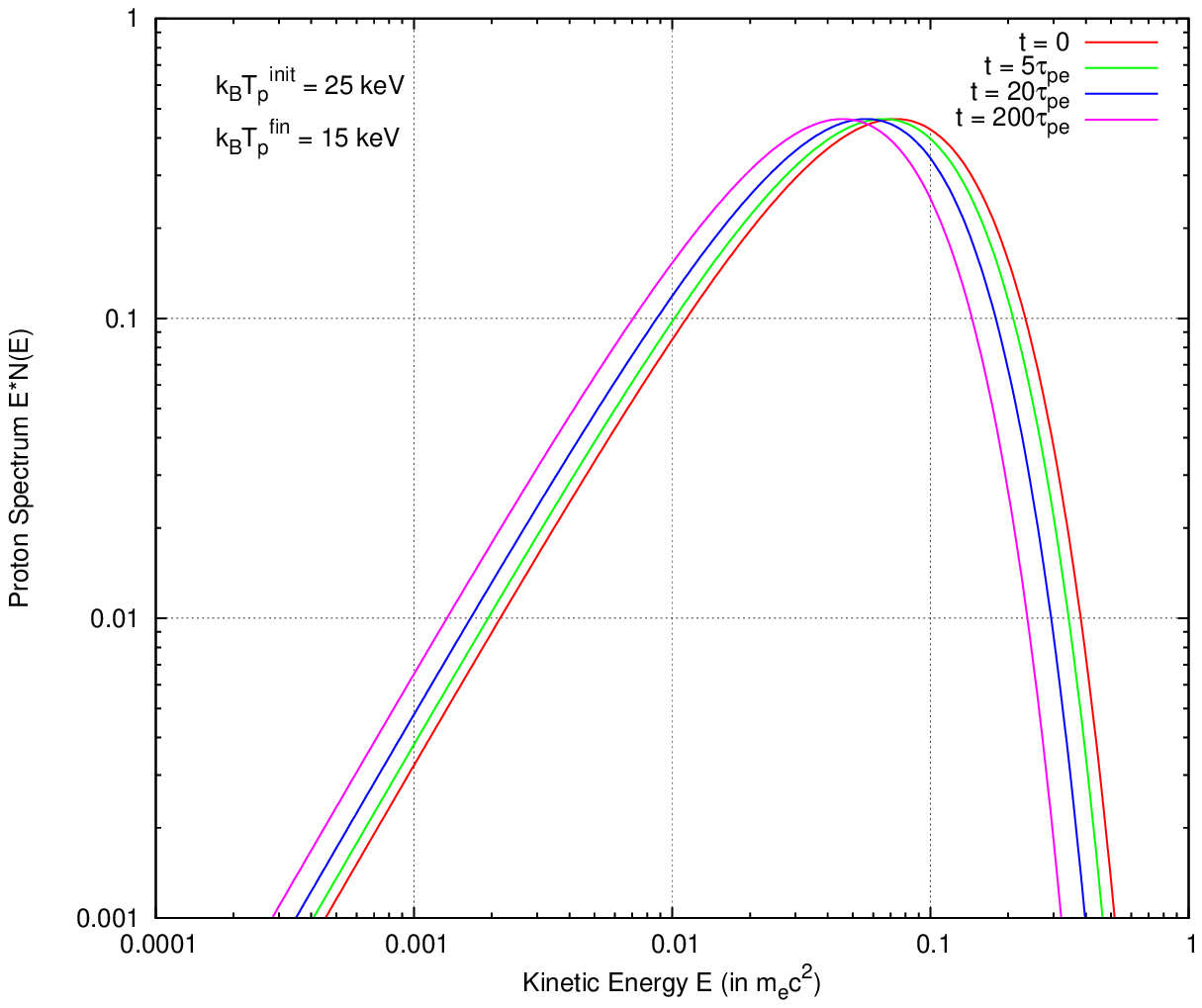}
\includegraphics[width=0.48\textwidth, height=0.3\textheight]{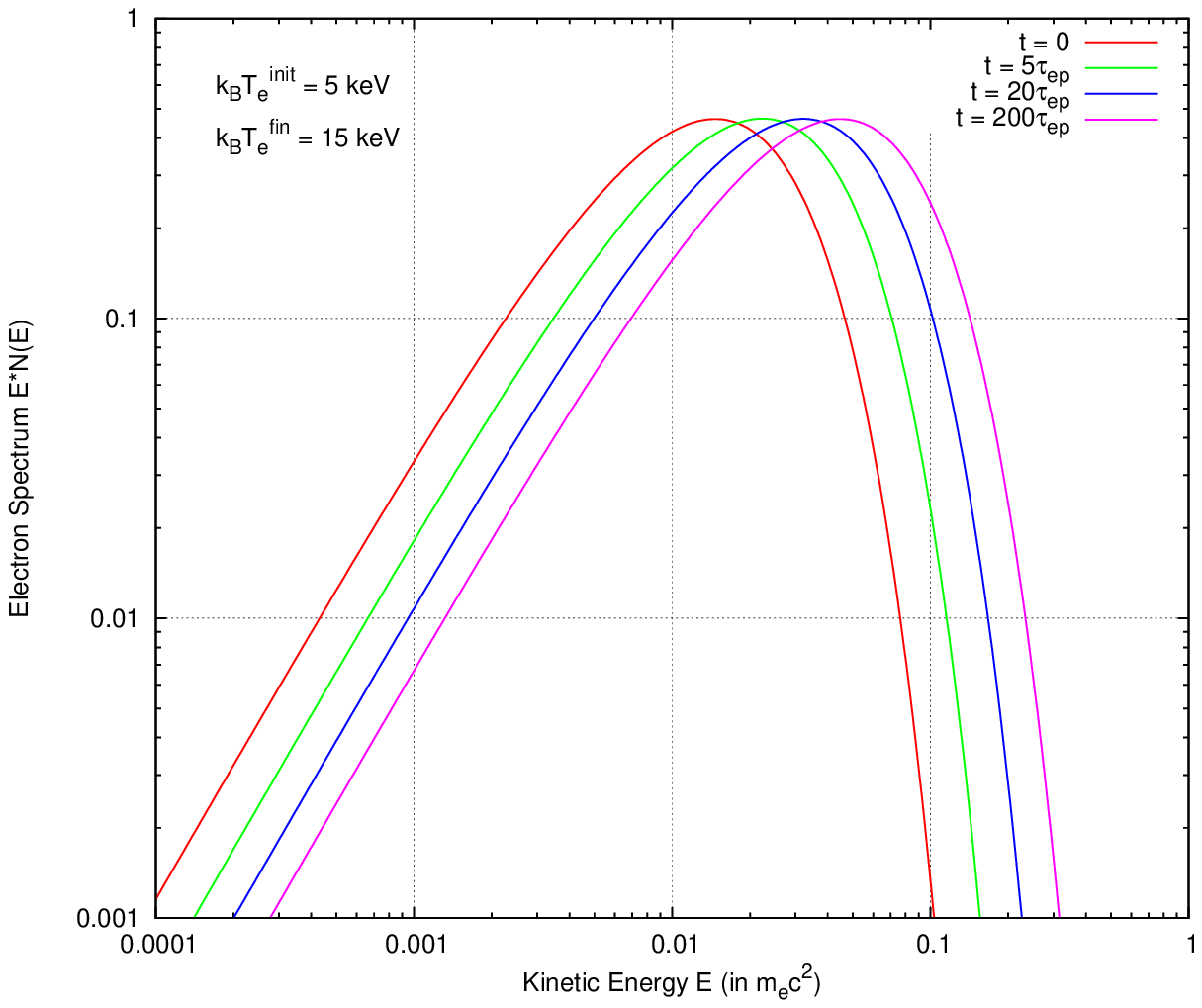}
\caption{Time evolution of distributions of protons (left) and electrons
(right) with initial Maxwellian distributions with $kT_p^{\rm init}=25$
keV and $kT_e^{\rm init}=5$ keV. Note that the distributions remain Maxwellian
and approach $kT^{\rm fin}=15$ keV after several hundred $\tau_{pe}$ and
$\tau_{ep}$, respectively, calculated at the initial
temperatures.} 
\label{fig:therm1}
\end{center}
\end{figure}

\begin{figure}[!ht]
\begin{center}
\includegraphics[width=0.48\textwidth, height=0.3\textheight]{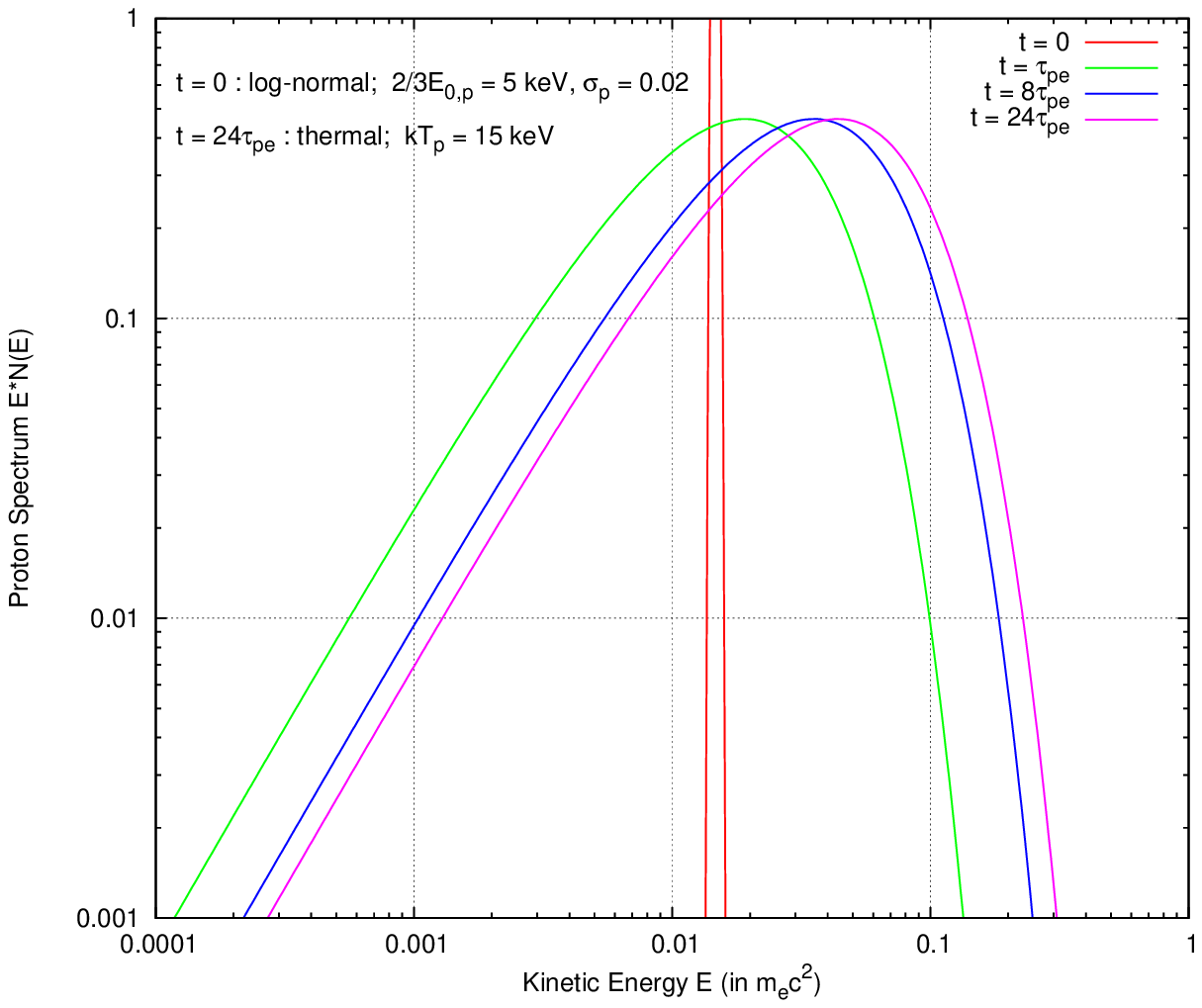}
\includegraphics[width=0.48\textwidth, height=0.3\textheight]{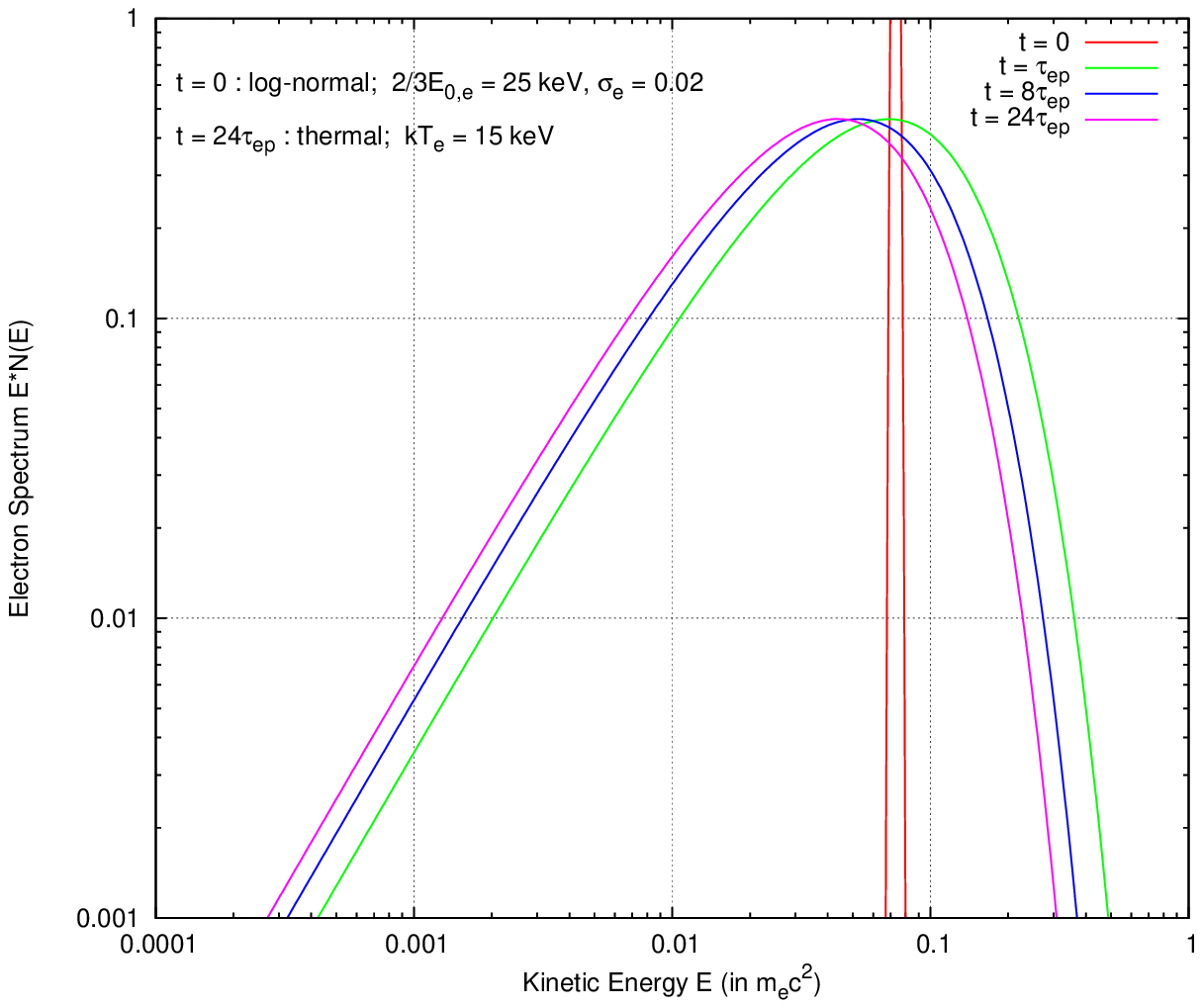}
\caption{Same as Figure \ref{fig:therm1} but  starting with more energetic
electrons with
initial log-normal distributions of $2/3E_{0,e}$ = 25 keV, $\sigma_p$ =
0.02, and protons with $2/3E_{0,e}$ = 5 keV. The electron (proton) distributions
approaches the Maxwellian shape
within
several $\tau_{\rm ee}$ ($\tau_{\rm pp}$) and is nearly an exact Maxwellian by
$t=\tau_{ep}$ ($t=\tau_{pe}$). Both distribution reach the equilibrium
temperature of 15 keV after 10's of $\tau_{ep}$ and $\tau_{pe}$, respectively.}
\label{fig:therm2}
\end{center}
\end{figure}

First, we consider the thermalization  when the protons and electrons
have different initial temperatures. Figure \ref{fig:therm1} shows the time
evolution of the proton and electron distributions with
initial temperatures  $kT_p$ = 25 keV and $kT_e$ = 5
keV.  As expected,  both species remain
Maxwellian throughout and their temperatures approach the average of the two 
initial temperatures over several $\tau_{pe}$'s and $\tau_{ep}$'s, respectively
calculated at the initial temperature.
We get similar results with initial values reversed; $kT_p$ =
5 keV and  $kT_e$ = 25 keV.%
\footnote{Figures showing this results, and several  others
not shown in this paper, can be found in Byungwoo Kang's (2013) Senior Thesis at
Stanford.}
Next, we consider the thermalization of the protons and electrons when they 
initially have a narrow log-normal distributions. The
normalized
log-normal distribution function is given by
\begin{equation}
f(E; E_0, \sigma) =
\frac{1}{E\sigma\sqrt{2\pi}}\mathrm{exp}\left(-\frac{\ln{E/E_0}}{2\sigma^2}
\right),
\end{equation}
whose mean is $E_0e^{\sigma^2/2}$. Note that, for $\sigma \ll 1$,
$E_0e^{\sigma^2/2}\approx E_0$, and the log-normal distribution becomes
essentially a Dirac delta function centered at $E_0$. In this
case, the proton and electron distributions should approach the Maxwellian shape
over several $\tau_{pp}$'s and $\tau_{ee}$'s respectively. At the same time,
their temperatures should also converge to $3E_0/2$ over several $\tau_{pe}$'s
and $\tau_{ep}$'s respectively. In Figure
\ref{fig:therm2} we show the time evolution of the proton and
electron distributions, when the protons and electrons are initially in
log-normal distributions with
$2/3E_{0,e}$ = 5 keV, $\sigma_p$ = 0.02, and $2/3E_{0,p}$ = 25 keV, $\sigma_e$ =
0.02, respectively. Not shown here, within 10's of $\tau_{ee}\ll \tau_{ep}$ and
$\tau_{pp}\ll
\tau_{ep}$, respectively, electrons and protons acquire Maxwellian distributions
(as can be seen in this figure this is the case at $t\sim\tau_{ep}$ and
$t\sim\tau_{pe}$,respectively. However it takes several tens of these timescale
before they both reach the final equilibrium of $kT_e=kT_p=15$ keV. Similar
results are obtained with initially more energetic protons. Similar results are
obtained starting with 25 keV protons and 5 keV electrons.

\section{Inelastic Energy Loss Rates and Times}

The inelastic energy loss rates for electrons are due to well known processes
of synchrotron, inverse Compton (with ${\dot E}\propto u\times
\beta_e^2\gamma_e^2$, where $u$ is the total energy density of magnetic field
and soft photons) and electron proton bremsstrahlung with  
\beq\label{ebrem}
{\dot E}^{\rm br}_{ep}=\left({4\alpha\over 3\pi \ln
\Lambda}\right){m_ec^2\over\tau_{\rm Coul}}\beta_e\gamma_e\chi(\gamma_e),
\eeq
where $\alpha=1/137$ is the fine structure constant and $\chi(\gamma)\simeq
a+b\ln \gamma$ is a slowly varying function.  At relativistic energies this rate
should
be increased by nearly an equal amount due to electron-electron bremsstrahlung.
Thus, bremsstrahlung loss becomes
dominant compared to elastic $e-e$ scattering at Lorentz factors $\geq 23\ln
\Lambda$. As shown in Figure \ref{hiEtime}, adding these rate modifies the
electron energy time scales shown in Figure \ref{ac-losst} at high energies.

For protons the main inelastic loss rate is due to hadronic interactions. For
$<30$ MeV protons the main loss is due to  the proton-ion  interactions
producing ionic de-excitation lines in
the 1-7 MeV range. Loss rate of  $>30$ to  $\sim 300$ MeV protons is dominated
by neutron production. Pion production starts at
threshold energy $E_{\rm th}\sim 300$ MeV and dominates all losses
(including Coulomb) above few
GeV.  
Proton-electron interactions can produce X- and gamma-rays via (inverse)
bremsstrahlung process. Since energy loss rate is an
invariant quantity it follows that ${\dot E}^{\rm br}_{pe}$ will be equal to
that of e-p bremsstrahlung in the rest frame of the proton, so that this rate
is also given by Eq.
(\ref{ebrem}) with electron velocity and Lorentz factor replaced with that of
protons. The inverse bremsstrahlung
loss  becomes negligible for  proton velocities less than the electron
thermal velocity, similar to the Coulomb $p-e$ loss (see Figure
\ref{ac-losst}).%

One can get a simple approximate expression for the energy loss rate due to
$p-p$ hadronic interactions at relativistic energies but there are no simple
furmulae for semi-relativistic energies and at energies near the threshold.
We combine  previously published results on these interaction   and  derive
a simple formulas  in three energy
regimes  with few percent accuracy at the boundaries of these regimes.

\subsection{$E_h=100 < E < 10^8$}

For this very high energy regime, we make use of the work of Kelner et al.
(2006) on the energy spectra of the secondaries particles produced by $p-p$
interactions. If we denote the number $dN_i$ of secondaries in the
kinetic energy
interval $(E_i,E_i+dE_i)$ produced by a
proton of energy $E_p$ (in unit of $m_pc^2$) by  $dN_i = F_i(E_i,E_p)dE_i$,
then the energy loss rate can be written as
\begin{equation}\label{edotGEN}
{\dot{E}}_{p-p} = \beta cn
{\sigma}_{inel}(E_p)\cdot\left(\displaystyle{\sum_{i}}
\int_0^{E_p} F_i(E_i,E_p)E_idE_i\right),
\end{equation}
where the index $i$ is summed over all  secondaries and ${\sigma}_{inel}(E_p)$
is the inelastic $p-p$ interaction cross section
that encodes information as to how many inelastic $p-p$ interactions occur (not
the number of the secondaries produced, as some cross sections defined
elsewhere do).

There are two choices as to counting the secondaries. We can count the
relatively stable secondaries, which are gamma-rays, electrons and 
electron and muon neutrinos (including their antiparticles). Alternatively, we
can
count the intermediate secondaries that decay
into the  above  mentioned more stable secondaries, which are pions and eta
mesons. The energy loss rate calculated in
both ways agree  with each other within few percent, and can be approximated
as 

\beq\label{edothi}
{\dot{E}}_{p-p}=C\times (E_p/E_h)^{1.09}, \,\,\,\,{\rm where} \,\,\,\,
C=1.26\times 10^{-24} c n m_pc^2 \beta_p.
\eeq

\subsection{$E_l=2.2 < E < E_h=100$}

For this intermediate energy regime, we adopt Stephens and Badhwar's model
(Stephens \& Badhwar 1981; see also Dermer 1986), where the pions are the main
particles produced and we get equal contributions from each pion to the sum in
Equation (\ref{edotGEN}) with
\begin{equation}\label{edotMID}
F_\pi(E_{\pi},E_{p})=\frac{2\pi\sqrt{E_{\pi}^2-m_{\pi}^2}c^4}{<\xi
\sigma_{\pi}(E)>}\int_{\cos{{\theta}_{max}}}^1
(E_\pi d^3\sigma/d{p_\pi}^3)d\cos{\theta},
\end{equation}
where $\cos{{\theta}_{max}}
=({\gamma}_{c}E_{\pi}-E_{max}^*)/({\beta}_c{\gamma}_c p_{\pi})$,
$E_{\rm max}^*=[E^2_{\rm tot}-(4m_p^2+m_{\pi}^2)c^4]/(2E_{\rm tot})$, and 
\begin{equation}\label{cross}
E_{\pi}\frac{d^3\sigma}{d{p_{\pi}}^3} = Af(E)(1-\tilde{x})^q
\mathrm{exp}[-Bp_{\perp}], 
\end{equation}
where $q=C_1+C_2p_\perp+c_3p_\perp^2$, with experimentally derived values of
$A = 140$, $B = 5.43$, $C_1 = 6.1$, $C_2 =
3.3$, and $C_3=0.6$.
In these equations 
$\tilde{x}^2=(p_{\parallel}/p^*_{\rm
max})^2+4(p_{\perp}^2c^2+m_{\pi}^2c^4)/E^2_{\rm tot}$;
all quantities measured in the c.m.s. 
Using these equations we find an energy loss rate 
${\dot{E}}_{p-p}=C\times [(E_p-E_l)/E_h + 0.012]$

\begin{figure}[!ht]
\begin{center}
\includegraphics[width=0.8\textwidth]{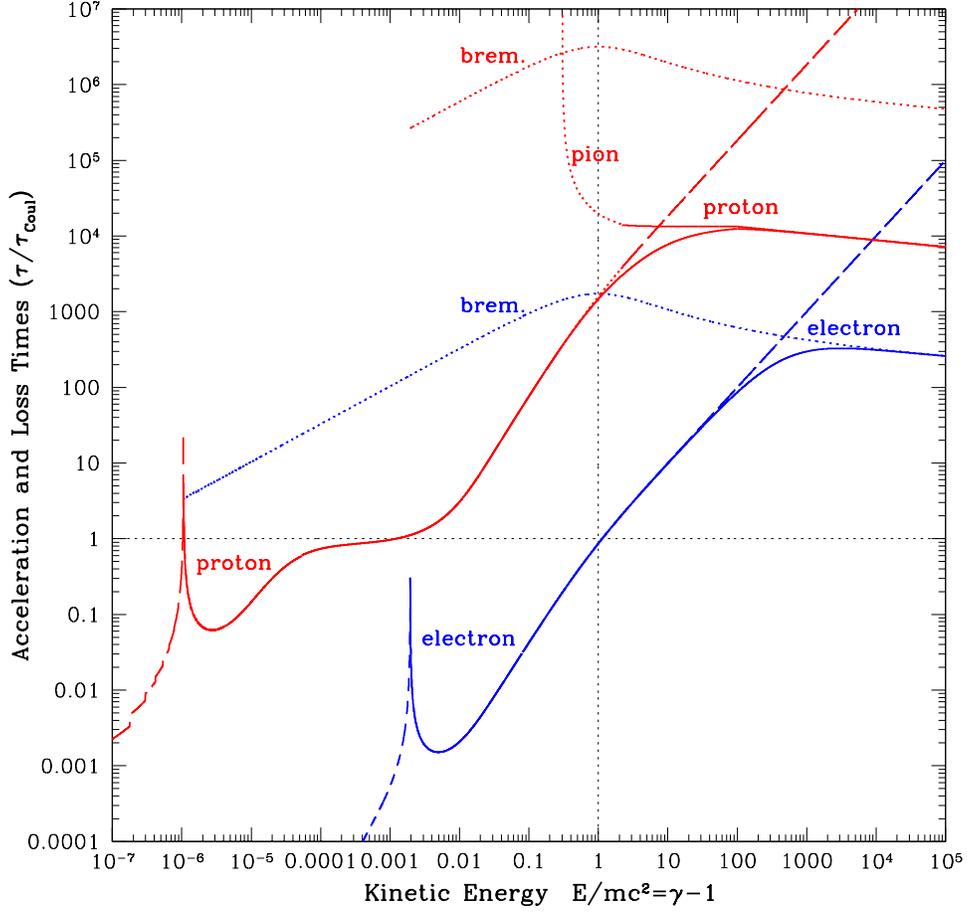}
\caption{Electron (lower, blue) and proton (upper, red) energy loss timescales
for elastic
(Coulomb with $\ln \Lambda=20$; dashed) and inelastic (dotted) e-p and p-e
bremsstrahlung  and p-p pion production processes. The solid lines give  total
energy loss timescale. As in Figure \ref{ac-losst} the
Coulomb loss timescale  includes both  warm-target e-e and p-p rates
(at temperature $kT = 2$ keV) and relativistic cold-target p-e rate.}
\label{hiEtime}
\end{center}
\end{figure}

\subsection{$E_{th}=0.3 < E < E_l=2.2$}

For the lowest energy range down to  the threshold kinetic energy we use the
Stecker's model (Stecker 1970; also see Dermer 1986), where we can approximate
the $\pi^0$ production
process by the so-called 'isobar-plus-fireball model'.%
\footnote{Stephens and Badhwar (1981) argue that their representation works
quite well in this energy  range too, despite the fact that the energy
range of
the experimental data used to determine these constants was around a few ten
GeV. However, Dermer (1986) shows that, near the threshold energy, Stecker's
model fits better with experimental pion energy spectra than Stephens-Badhwar
model.} 
In this model, we will
assume that all pions are produced in two ways: via intermediate production and
decay of the $\Delta(1.238)$ non-strange isobar, and from a thermal pion gas
created from the  remaining available energy in the c.m.s of the collision where
the pions are given an energy distribution very similar to a Maxwell-Boltzmann
type distribution. In particular, for $E < 2.2$, we can assume that all pion
production occurs through the isobar production mode because it is the most
dominant mode in this energy regime (the fireball mode
becomes
dominant when the total energy of the incident proton is greater than 5 GeV). We
further assume that the isobars of mass $m_{\Delta}$ carry momentum either
directly forward or directly backward in the c.m.s, and that the low-energy
isobar
production process produces pions through the two-stage decay
\begin{align*}
p + p \rightarrow p + & \Delta(1.238)\\
& \downarrow \\
& p +\pi^0.
\end{align*}

The normalized production spectrum of pions, to be used in Equation
(\ref{edotGEN}), in this model is given
by
\begin{equation}
F_\pi(E_{\pi},E_p)=w_r(E_p)\int_{m_p+m_{\pi^0}}^{s^{1/2}-m_p}dm_{
\Delta}B(m_{\Delta})f(E_{\pi},E,m_{\Delta}),
\end{equation}
where 
\begin{align}
f(E_{\pi},E,m_{\Delta}) & = \frac{1}{2m_{\Delta}}
\{(2\beta_{\Delta}^{+}\gamma_{\Delta}^{+}\beta_{\pi}^{'}\beta_{\pi}^{'})^{-1}H[
\gamma_{\pi};\gamma_{\Delta}^{+}\gamma_{\pi}^{'}(1-\beta_{\Delta}^{+}\beta_{\pi}
^{'}),\gamma_{\Delta}^{+}\gamma_{\pi}^{'}(1+\beta_{\Delta}^{+}\beta_{\pi}^{'})]
\nonumber \\ 
& +
(2\beta_{\Delta}^{-}\gamma_{\Delta}^{-}\beta_{\pi}^{'}\beta_{\pi}^{'})^{-1}H[
\gamma_{\pi};\gamma_{\Delta}^{-}\gamma_{\pi}^{'}(1-\beta_{\Delta}^{-}\beta_{\pi}
^{'}),\gamma_{\Delta}^{-}\gamma_{\pi}^{'}(1+\beta_{\Delta}^{-}\beta_{\pi}^{'})]
\},
\end{align}
and the normalized isobar mass spectrum is given by the Breit-Wigner
distribution
$B(m_{\Delta})=(\Gamma/\pi)(1/(m_{\Delta}-m_{\Delta}^0)^2+{\Gamma}^2$ 
with the normalization factor
$w_r(E)=\pi[\arctan(s^{1/2}-m_p-m_{\Delta}^0/\Gamma)-\arctan(m_p+m_{\pi^0}-m_{
\Delta}^0/\Gamma)]^{-1}$.
However, since $\Gamma$ is sufficiently small, we
can approximate $B(m_{\Delta})$ by the Dirac-delta function
$\delta(m_{\Delta}-m_{\Delta}^0)$ and therefore ignore the normalization factor
$w_r(E)$.

Here $H[x;a,b]=1$ if $a \le x \le b$ and 0 otherwise, and $\gamma_{\pi}$ is the
pion Lorentz factor in the lab system (LS). The Lorentz factors of the forward
(+) and backward (-) moving isobars  are $\gamma_{\Delta}^{\pm} =
\gamma_c\gamma_{\Delta}^{\ast}(1\pm\beta_c\beta_{\Delta}^{\ast})$, where
$\gamma_c=s^{1/2}/2m_{p}$ is the Lorentz factor of the center of mass (CM) with
respect to the LS, and
$\gamma_{\Delta}^{\ast}=(s+m_{\Delta}^2-m_{\pi}^2)/2s^{1/2}m_{\Delta}$ is the
Lorentz factor of the isobar in the CM. The pion Lorentz factor in the rest
frame of the $\Delta$-isobar is
$\gamma_{\pi}^{'}=(m_{\Delta}^2+m_{\pi}^2-m_p^2)/2m_{\Delta}m_{\pi}$.

Again with the help of these equations and Equation (\ref{edotGEN}) we obtain an
energy loss rate of $C\times[5.9\times 10^{-3}(E_p-E_{th}) + 2.2\times
10^{-4}(E_p-E_{th})^2]$.

Now putting all this together we obtain energy loss rate as a
function of proton energy $E$ (in units of $m_pc^2$) as
\begin{multline}
{\dot{E}}_{p-p}=\frac{1.26m_pc^2\beta}{\tau_{\rm Coul}\ln \Lambda}	
\begin{cases}
0.011(E-E_{\mathrm{th}})+2.2\times 10^{-4}(E-E_{\mathrm{th}})^2, &
E_{\mathrm{th}} \le E \le
E_l, \\
(E-E_l)/E_h+0.022, & E_l \le E \le E_h, \\
(E/E_h)^{1.09}, & E_h \le E \le 10^8,
\end{cases}
\end{multline}
where $E_{\mathrm{th}} = 0.3$, $E_l = 2.2$ and $E_h = 100$.
In Figure \ref{hiEtime} we show energy loss for both electrons and protons
extending it to higher energies than those shown in Figure \ref{ac-losst}.

%Inside the square bracket of \bibitem, "a, 1" is added to avoid conflict with the new version of natbib.

\end{document}